\documentclass{nature}
\usepackage{amssymb}
\usepackage{aas_macros}
\usepackage{graphicx}
\usepackage{pdflscape}
\usepackage{hyperref}
\usepackage{txfonts}
\usepackage{textcomp}
\usepackage{threeparttable}
\usepackage{marvosym}

\usepackage{amssymb,amsmath}
\usepackage{xcolor}
\usepackage{ulem}
\usepackage{pdfpages}
\usepackage[caption = false]{subfig}
\usepackage{caption}
\makeatletter
\usepackage[symbol]{footmisc}
\newcommand{\EXTTAB}[1] {Supplementary Table}
\usepackage{longtable}
\usepackage{tabularx}

\let\saved@includegraphics\includegraphics
\AtBeginDocument{\let\includegraphics\saved@includegraphics}
\renewenvironment*{figure}{\@float{figure}}{\end@float}
\makeatother
\title{Revealing the Temporally Stable Bimodal Energy Distribution of FRB 20121102A with a Tripled Burst Set from AI Detections}
\begin{document}

\maketitle
\author{
Y.~D. Wang$^{1,3}$\href{https://orcid.org/0000-0002-7372-4160}{\includegraphics[scale=0.08]{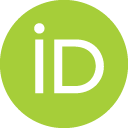}}\footnotemark[1],
J. Han$^{2}$\href{https://orcid.org/0000-0002-3717-0199}{\includegraphics[scale=0.08]{ORCIDiD.png}}\footnotemark[1],
P. Wang$^{1,4}$\href{https://orcid.org/0000-0002-3386-7159}{\includegraphics[scale=0.08]{ORCIDiD.png}}\footnotemark[1] \textsuperscript{\Letter},
D. Li$^{5,1}$\href{https://orcid.org/0000-0003-3010-7661}{\includegraphics[scale=0.08]{ORCIDiD.png}\textsuperscript{\Letter}},
H.~T Chen$^{6}$\href{https://orcid.org/0009-0004-6709-386X}{\includegraphics[scale=0.08]{ORCIDiD.png}\textsuperscript{\Letter}},
Y.~C. Tian$^{7}$\href{https://orcid.org/0009-0000-3599-4605}{\includegraphics[scale=0.08]{ORCIDiD.png}},
E.
G\"{u}gercino\u{g}lu$^{1}$,
J.~N. Tang$^{8}$, 
Z.~H. Zhang$^{8}$,
K.~C. Wu$^{8}$,%
X.~L. Zhang$^{8}$,%
Y.~H. Zhu$^{1,3}$\href{https://orcid.org/0009-0009-8320-1484}{\includegraphics[scale=0.08]{ORCIDiD.png}},
J.~H. Cao$^{1,3}$\href{https://orcid.org/0009-0000-7501-2215}{\includegraphics[scale=0.08]{ORCIDiD.png}},
M.~T. Chen$^{9}$\href{https://orcid.org/0009-0008-7635-2949}{\includegraphics[scale=0.08]{ORCIDiD.png}},
J.~P. Feng$^{10}$\href{https://orcid.org/0009-0008-6783-5762}{\includegraphics[scale=0.08]{ORCIDiD.png}},
Z.~Y. Huai$^{11}$\href{https://orcid.org/0009-0009-1219-5128}{\includegraphics[scale=0.08]{ORCIDiD.png}},
Z.~T. Lin$^{12}$\href{https://orcid.org/0000-0001-5695-8734}{\includegraphics[scale=0.08]{ORCIDiD.png}},
J.~M. Luan$^{13}$,
H.~B. Wang$^{14}$\href{https://orcid.org/}{\includegraphics[scale=0.08]{ORCIDiD.png}},
J.~J. Zhao$^{15,16}$\href{https://orcid.org/0009-0004-6675-5446}{\includegraphics[scale=0.08]{ORCIDiD.png}},
\\
C.~W. Tsai$^{1,3,4}$\href{https://orcid.org/0000-0002-9390-9672}{\includegraphics[scale=0.08]{ORCIDiD.png}},
W.~W. Zhu$^{1,4}$\href{https://orcid.org/0000-0001-5105-4058}{\includegraphics[scale=0.08]{ORCIDiD.png}},
Y.~K. Zhang$^{1,3}$\href{https://orcid.org/0000-0002-8744-3546}{\includegraphics[scale=0.08]{ORCIDiD.png}},
Y. Feng$^{17,18}$\href{https://orcid.org/0000-0002-0475-7479}{\includegraphics[scale=0.08]{ORCIDiD.png}},
A.~Y. Yang$^{1}$\href{https://orcid.org/0000-0003-4546-2623}{\includegraphics[scale=0.08]{ORCIDiD.png}},
D.~K. Zhou$^{17}$\href{https://orcid.org/0000-0002-7420-9988}{\includegraphics[scale=0.08]{ORCIDiD.png}},
J.~H. Fang$^{17}$\href{https://orcid.org/0000-0001-9956-6298}{\includegraphics[scale=0.08]{ORCIDiD.png}},
J.~Y. Xu$^{17}$,
C.~H. Niu$^{19}$\href{https://orcid.org/0000-0001-6651-7799}{\includegraphics[scale=0.08]{ORCIDiD.png}},
J.~R. Niu$^{1,3}$\href{https://orcid.org/0000-0001-8065-4191}{\includegraphics[scale=0.08]{ORCIDiD.png}},
J.~M. Yao$^{20}$,
C.~F. Zhang$^{1,21}$\href{https://orcid.org/0000-0002-4327-711X}{\includegraphics[scale=0.08]{ORCIDiD.png}},
R.~S. Zhao$^{22}$\href{https://orcid.org/0000-0002-1243-0476}{\includegraphics[scale=0.08]{ORCIDiD.png}},
L. Zhang$^{1,23}$\href{https://orcid.org/0000-0001-8539-4237}{\includegraphics[scale=0.08]{ORCIDiD.png}},
J.~S. Zhang$^{1,3}$\href{https://orcid.org/0009-0005-8586-3001}{\includegraphics[scale=0.08]{ORCIDiD.png}},
W.~J. Lu$^{1,3}$\href{https://orcid.org/0000-0001-5653-3787}{\includegraphics[scale=0.08]{ORCIDiD.png}},
Q.~Y. Qu$^{3}$\href{https://orcid.org/0009-0005-5413-7664}{\includegraphics[scale=0.08]{ORCIDiD.png}}
}
\makeatletter
\def\thanks#1{\protected@xdef\@thanks{\@thanks
        \protect\footnotetext{#1}}}
\makeatother

\maketitle
\footnotetext[1]{These authors contributed equally to this work.}
\footnotetext{\Letter~ wangpei@nao.cas.cn; dili@mail.tsinghua.edu.cn; chenhanting@huawei.com}
\begin{affiliations}
\item National Astronomical Observatories, Chinese Academy of Sciences, Beijing 100101, China
\item School of Artificial Intelligence, Beijing University of Posts and Telecommunications, Beijing 100876, China
\item University of Chinese Academy of Sciences, Beijing 100049, China
\item Institute for Frontiers in Astronomy and Astrophysics, Beijing Normal University,  Beijing 102206, China
\item New Cornerstone Science Laboratory, Department of Astronomy, Tsinghua University, Beijing 100084, China
\item Huawei Noah's Ark Lab, Beijing 100085, China
\item School of Intelligence Science and Technology, Peking University, Beijing 100871, China
\item Computer Network Information Center, Chinese Academy of Sciences, Beijing 100083, China
\item Department of Chemistry, National University of Singapore, Singapore 117549, Singapore
\item Department of Astronomy, University of Science and Technology of China, Hefei 230026, China
\item California Institute of Technology, California, CA 91125, USA 
\item Department of Astronomy, Tsinghua University, Beijing 100084, China
\item Rice University, Houston, Texas, 77005, USA
\item Nagqu Senior High School, Nagqu 852000, China
\item Department of Astronomy, School of Physics and Astronomy, Shanghai Jiao Tong University, Shanghai 200240, China 
\item Tsung-Dao Lee Institute, and Shanghai Key Laboratory for Particle Physics and Cosmology, Shanghai Jiao Tong University, Shanghai 200240, China
\item Zhejiang Lab, Hangzhou, Zhejiang 311121, China
\item Institute for Astronomy, School of Physics, Zhejiang University, Hangzhou 310027, China
\item Institute of Astrophysics, Central China Normal University, Wuhan 430079, China
\item Xinjiang Astronomical Observatory, Chinese Academy of Sciences, Urumqi 830011, China
\item Department of Astronomy, Peking University, Beijing 100871, China
\item Guizhou Provincial Key Laboratory of Radio Astronomy and Data Processing, Guizhou Normal University, Guiyang 550001, China
\item Centre for Astrophysics and Supercomputing, Swinburne University of Technology, P.O. Box 218, Hawthorn, VIC 3122, Australia

\end{affiliations}
\vspace{0.15in}
\begin{abstract}

\end{abstract}

\linespread{1.15}

\noindent \textbf{Active repeating Fast Radio Bursts (FRBs), with their large number of bursts, burst energy distribution, and their potential energy evolution, offer critical insights into the FRBs emission mechanisms. Traditional pipelines search for bursts through conducting dedispersion trials and looking for signals above certain fluence thresholds, both of which could result in missing weak and narrow-band bursts. In order to improve the completeness of the burst set, we develop an End-to-end DedispersE-agnostic Nonparametric AI model (EDEN), which directly detect bursts from dynamic spectrum and is the first detection pipeline that operates without attempting dedispersion. We apply EDEN to archival FAST L-band observations during the extreme active phase of the repeating source FRB 20121102A, resulting in the largest burst set for any FRB to date, which contains 5,927 individual bursts, tripling the original burst set. The much enhanced completeness enables a refined analysis of the temporal behavior of energy distribution, revealing that the bimodal energy distribution remains stable over time. It is rather an intrinsic feature of the emission mechanisms than a consequence of co-evolving with burst rate.}

The completeness of active repeating FRBs burst set is crucial, as the burst set can enable an in-depth investigation of the burst energy distribution and its temporal behavior, providing important insights into the underlying FRBs emission mechanism. Existing search pipelines, limited by fluence threshold, can result in incomplete detection of weak and narrow-band bursts. In this work, we develop an End-to-end DedispersE-agnostic Nonparametric AI model (EDEN) that provides a much more complete set of detected bursts. The proposed sensitive algorithm is fully end-to-end, processing dynamic spectra from telescopes and directly outputting detection results. This streamlined, end-to-end design allows the algorithm to operate more faster than existing grid-search-based algorithms.

\begin{figure*}
    \centering
    \includegraphics[width=\textwidth]{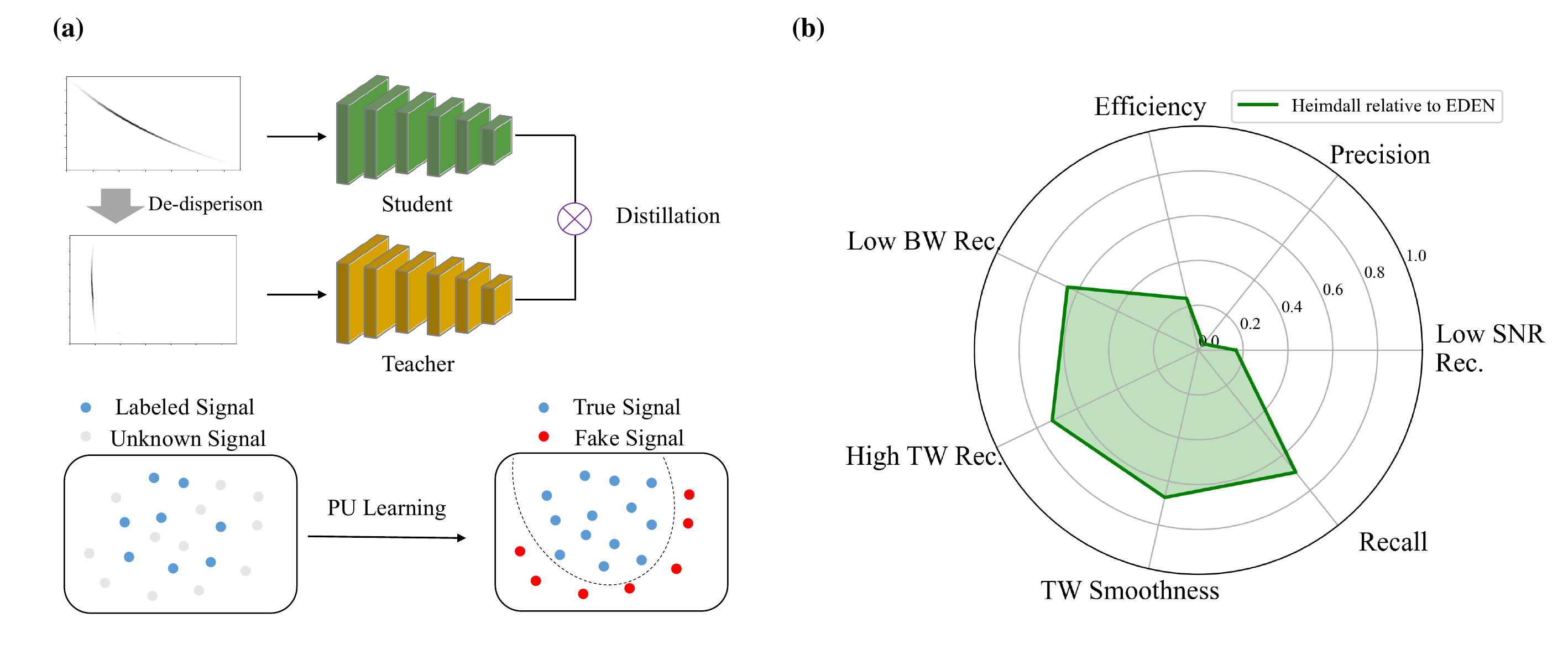}
    \caption{{Overview of the proposed model EDEN for FRB detection.} Panel (a): The pipeline for training the model, which incorporates Teacher-Student Learning and Positive-Unlabeled Learning to enhance the diversity of detectable signals. Panel (b): A comparison between EDEN and the conventional Heimdall method for FRB detection. EDEN outperforms Heimdall in all aspects, particularly in terms of speed, precision, and the detection of high-time width (TW) and low-bandwidth (BW) signals.}
    \label{fig:ai_versus_heimdall}
\end{figure*}

To obtain a more complete burst set that encompasses corner cases, e.g. faint or band limited bursts, we employ several techniques to train EDEN for enhanced completeness. We adopt a Teacher-Student learning approach, where the student learns from a diverse set of simulated signals, ensuring that a sufficient number of weak burst signals are included, thereby improving weak burst detection. Additionally, we leverage Positive-Unlabeled (PU) training to address the challenge of weak bursts potentially remaining unidentified in the training dataset. The schematic diagram of EDEN is shown in Figure~\ref{fig:ai_versus_heimdall} Panel (a). 

To thoroughly evaluate the advantages of EDEN over the traditional Heimdall~\cite{Heimdall}/PRESTO\cite{ransom2001new} liked method, we conducted extensive signal simulations by injecting numerous synthetic signals into a real FAST noise background. A total of 10,000 signal samples were generated (see Methods). A quantitative comparison of EDEN and the conventional Heimdall algorithm across seven metrics — Efficiency, overall Precision, Recall, Recall for Low Bandwidth, Recall for High Time Width, Time Width smoothness, and Recall for Low SNR is shown in Figure~\ref{fig:ai_versus_heimdall} Panel (b).

Efficiency is significantly improved due to the end-to-end nature of the AI algorithm, which detects signals more than four times faster than Heimdall. EDEN is also capable of identifying signal patterns that differ from those detected by Heimdall, with particular sensitivity to faint signals. Its ability to detect low bandwidth and high burst width signals has notably surpassed that of Heimdall. The result of Precision, measuring the proportion of true signals among those identified by the model, shows that EDEN achieves more than 29 times the precision of Heimdall.

Using the EDEN algorithm, we conducted a deep search on the FRB 20121102A dataset reported in Ref.~\cite{Li_2021}. The 2019 observations of FRB 20121102A primarily focused on bright bursts, whereas this work emphasizes the systematic search for weak bursts and tests the bimodal burst energy distribution at lower detection thresholds. We have detected a total of 5,927 independent bursts, tripling the number of detections. To calculate the total isotropic equivalent energy, we adopt Equation (9) in Ref.~\cite{Zhang_2018} to maintain consistency with Ref.~\cite{Li_2021} (see Methods) and yields a total energy of \(5.94 \times 10^{41}\) erg, nearly doubling the total energy of \(3.41 \times 10^{41}\) erg reported in Ref.~\cite{Li_2021}. 

\begin{figure*}
    \centering
    \includegraphics[scale=0.55]{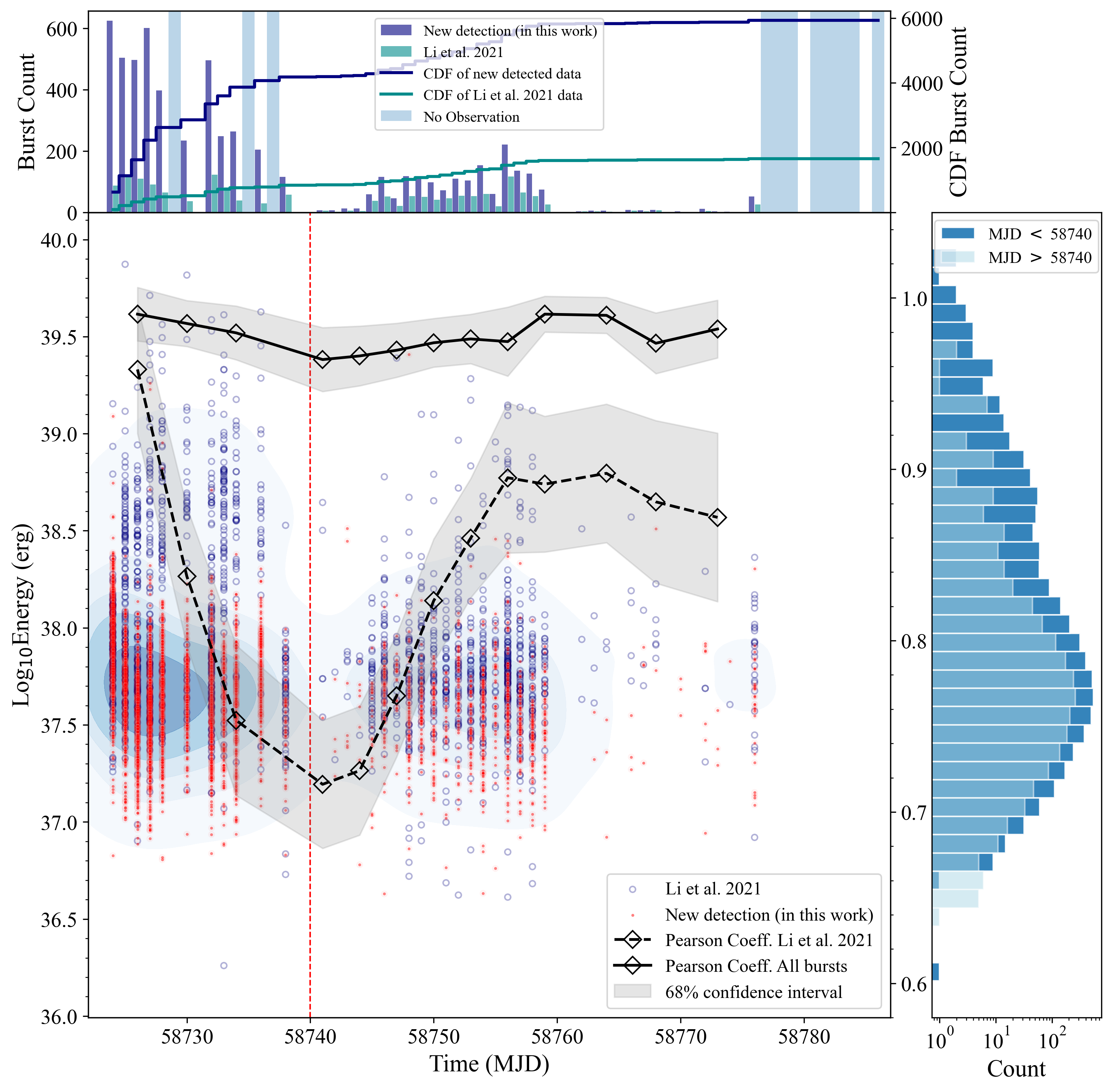}
    \caption{{Temporal energy distribution of the bursts.} The top panel compares burst counts between the newly detected bursts and the previously detected 1652 bursts. The burst count for each epoch and the cumulative burst count for the two burst sets are shown separately in deep blue and green, with light blue-shaded regions indicating periods without observation. The middle panel displays the burst energy distribution, where blue dots represent the energy of the original 1652 bursts, red dots represent the energy of the newly detected bursts in each observing session, and the blue contour represents the two-dimensional KDE of the burst distribution. Pearson correlation coefficient analysis between subsets of bursts is shown with black lines and with shaded area being the 68\% confidence interval. The vertical dashed line indicates the MJD at which the slope of the PCC curve is zero, marking the point where the datasets on either side exhibit the largest morphological difference. The right panel presents a histogram of isotropic burst energy, distinguishing bursts detected before (dark blue) and after (light blue) MJD 58740.}
    \label{fig-method-timeE}
\end{figure*}

This algorithm significantly contributed to forming a much more complete burst set for the 2019 observations of FRB 20121102A. Figure \ref{fig-method-timeE} illustrates the time-energy distribution of the bursts, with the comparison of the cumulative number of bursts detected in each epoch shown in the upper panel. The burst rate peaked at 495 hr$^{-1}$ during a one-hour observation on September 7th, four times the rate of the previous burst set. The lower-left panel of Figure \ref{fig-method-timeE} shows the distribution of burst energy in each epoch, with newly detected bursts in red and the previously detected 1,652 bursts in blue. This demonstrates that the newly detected bursts are concentrated in the lower energy range, filling the gap in previous detections.

To further investigate the temporal behavior of the burst energy, we conducted a quantitative correlation analysis of the morphological differences to verify the validity of the temporal evolution. We computed the Pearson correlation coefficient (PCC) and present in Figure \ref{fig-method-timeE}. The bursts in a given MJD, including those detected only in that epoch, are recorded in a subset \( b_i \), where \( i = 1, \dots, 41 \). The sets are defined as follows:
\begin{equation}
\begin{aligned}
    B_j &= \sum_{i=1}^{i=3+3j} b_i,~j = 0,1, \dots, 12 \\
    B_{\text{res}_j} &= \sum_{i=1}^{i=41} b_i - B_j,~j = 0,1, \dots, 12
\end{aligned}
\end{equation}
Starting from \( B_0 \), the dataset \( B_j \) is progressively enlarged by including subsequent \( b_i \) in steps of 3. The residual set \( B_{\text{res}_j} \) is obtained by subtracting \( B_j \) from the total set. We then normalize the sets using the estimated density function derived from the kernel density estimate (KDE). Specifically, the KDE is computed using a Gaussian kernel, which creates a continuous estimate of the probability density function based on the data. The correlation analysis is performed between the normalized sets \( B_j^{'} \) and \( B_{\text{res}_j}^{'} \), yielding the PCC. 

As depicted in Figure \ref{fig-method-timeE}, the black lines with diamond markers show the two trends of PCC. The vertical dashed line marks the MJD where the slope of the PCC curve is zero, indicating the point where the PCC reaches its minimum and the morphological difference between the datasets of two sides is maximal. Although a transition point around MJD 58740 is obvious for the previous burst set, it is not significant for the new burst set derived by our algorithm EDEN. It is noticeable that the trend of PCCs from all detected 5,927 bursts is stable and the values are close to 1. We employ bootstrap resampling with 1,000 iterations to derive the 68\% confidence interval (shaded regions in Figure \ref{fig-method-timeE}). We find from the new burst set that the variation of PCC values does not exceed the 1\(\sigma\) confidence level, indicating the energy distribution remains stable statistically over time.

We then tested the bimodal characteristic on the new burst set (see Methods). The bimodality, which is now commonly observed in active repeating FRBs such as FRB 20201124A~\cite{Xu_2022,ZhangYK_2022} and FRB 20220912A~\cite{ZhangYK_2023}, still exists in the tripled burst set. Such complexity of the energy spectrum shows that repeating FRBs can have bursts with diverse types, suggesting that different emission mechanisms may account for two peaks~\cite{PetroffDL_2022}. 

Notably, our Pearson analysis reveals that the bimodality persists throughout the entire observation period, demonstrating a statistical behavior distinct from previous understanding. Active FRBs exhibit variations in burst rate even during their active phases. Although the traditional pipeline previously identified a burst set of 1,652 bursts and was relatively large at that time, we could not determine whether the bimodal energy distribution only occurred during  particularly active epochs. This is because a larger proportion of bright bursts were detected during highly active epochs, leading to the derivation of two peaks. Thus, previous analysis has shown temporal evolution and suggested that the bimodal energy distribution co-evolves with burst rate. Now, with a large number of newly detected bursts, the burst rates have increased, and their variations still exist. Meanwhile, our more complete burst set is unaffected by burst rate variations. While burst rate continues to fluctuate, the temporal behavior of the bimodal energy distribution remains statistically stable, indicating the lack of temporal evolution. The temporal stability of the bimodal energy distribution further implies the possible existence of multiple stable emission mechanisms.

\begin{figure*}
	\centering
	\includegraphics[width=1\linewidth]{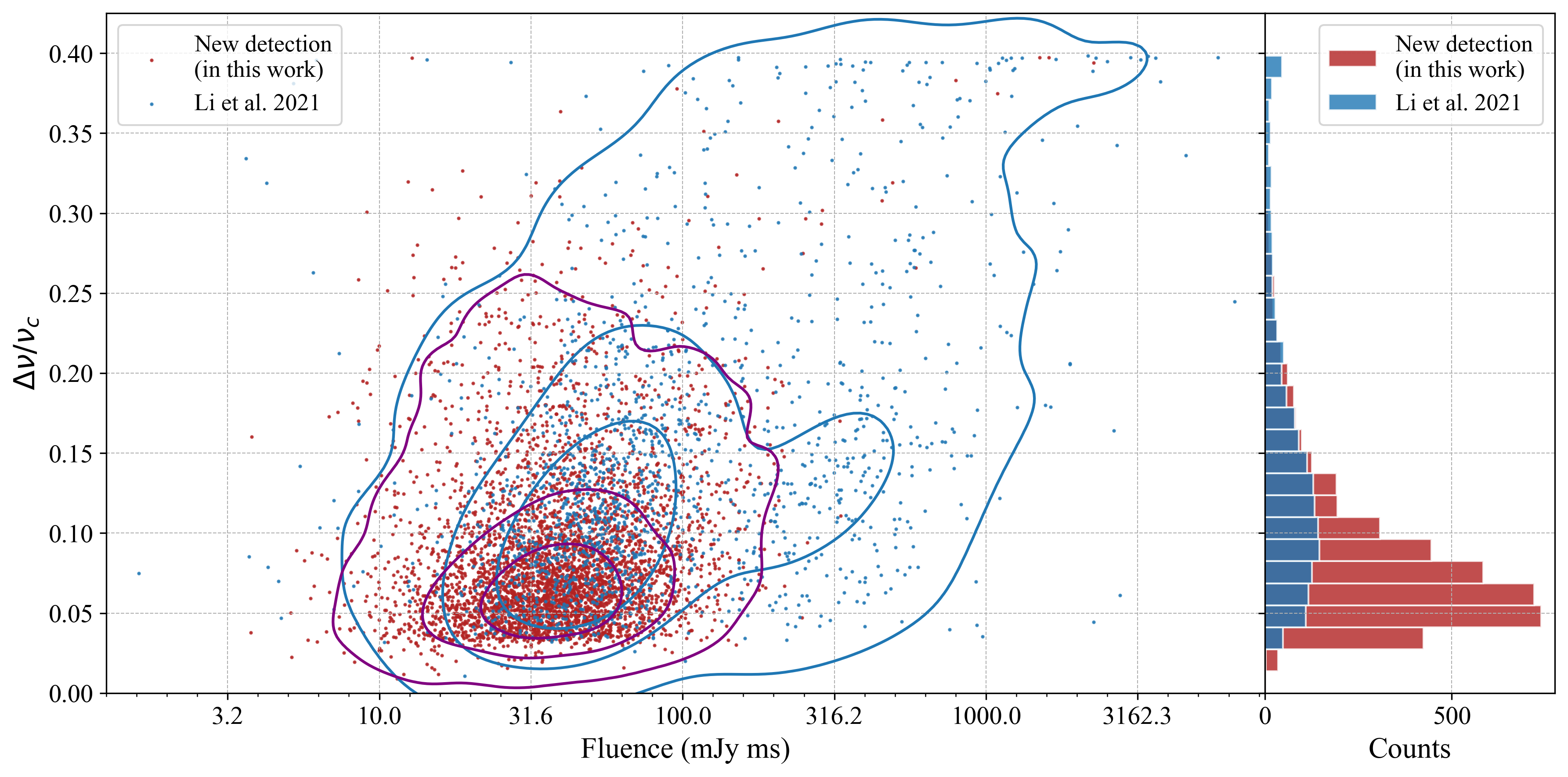}
	\caption{Distribution of the bandwidth-to-central-frequency ratio and fluence for the bursts. In the left panel, new detections are represented by red dots, while the detections reported in Ref.~\cite{Li_2021} are represented by blue dots. Three contour lines for each burst set partition the probability mass function into four regions. The density contours reveal two distinct clusters: one in the lower-left corner, predominantly consisting of new detections, characterized by lower fluence and narrower bandwidths, and another in the upper-right region, primarily composed of previously reported bursts, characterized by higher fluence and broader bandwidths. The right panel presents histograms for the two burst sets. Both panels show clustering in the narrow-bandwidth region, emphasizing the morphological differences between the two sets of detected bursts.}
	\label{fig-main-bw-ctrf}
\end{figure*}

In addition to identifying faint bursts, EDEN also excels in detecting narrow-bandwidth bursts. Figure~\ref{fig-main-bw-ctrf} shows the distribution of the ratio of bandwidth (\(\Delta\nu\)) to central frequency (\(\nu_c\)) versus fluence. The red markers represent new detections, which are clustered in the lower-left region of the parameter space, while the blue markers correspond to the original bursts, predominantly occupying the upper-right region. The contour distribution reveals two distinct clusters: one consisting of narrow-bandwidth, low-energy bursts, and the other of broad-bandwidth, high-energy bursts. This clustering indicates that EDEN primarily identifies narrow-bandwidth bursts in the new detections. The right panel further illustrates that the newly detected bursts are concentrated in a narrower bandwidth, consistent with the bandwidth distribution (see Methods) and highlighting the algorithm's superiority. Traditional pipelines, which require the calculation of the SNR across the entire bandwidth, inadvertently amplify the noise background for narrow-bandwidth bursts, thereby compromising detection performance.

We leveraged EDEN to successfully identify a large number of faint and narrow bandwidth bursts, tripling the original sample size. Previous studies had underestimated the total radiated energy, and the newly detected bursts have doubled the total isotropic equivalent energy. The much more complete burst set provides new insights into the temporal behavior of bimodal energy distribution, showing a statistically robust stability over time. It is likely to be an intrinsic feature of emission mechanisms instead of a product from co-evolution with burst rate, suggesting the likely existence of multiple stable emission mechanisms.

\bibliographystyle{naturemag}

\begin{addendum}
\item  This work is supported by National Natural Science Foundation of China (NSFC) Programs (No. 12588202, U1731238); by CAS International Partnership Program (No. 114-A11KYSB2016\\0008); by CAS Strategic Priority Research Program (No. XDB23000000); and the National Key R\&D Program of China (No. 2017YFA0402600); and the National SKA Program of China (No. 2020SKA0120200, 2022SKA0130100). D.L. is a New Cornerstone investigator. P.W. acknowledges support from the CAS Youth Interdisciplinary Team, the Youth Innovation Promotion Association CAS (id. 2021055), and the Cultivation Project for FAST Scientific Payoff and Research Achievement of CAMS-CAS. This work made use of the data from FAST (Five-hundred-meter Aperture Spherical radio Telescope) (https://cstr.cn/31116.02.FA\\ST). FAST is a Chinese national mega-science facility, operated by National Astronomical Observatories, Chinese Academy of Sciences.

\item[Author Contributions] 
Y.D.W. and P.W. led the data processing and analysis, J.H., H.T.C. and Y.C.T. led the algorithm development.
D.L. and H.T.C. are the conveners of the project, coordinated the science team, and launched the collaboration. 
Y.H.Z., J.H.C., J.N.T., Z.H.Z, K.C.W. and X.L.Z. provided equipment and technical support. 
M.T.C., J.P.F., Z.Y.H., Z.T.L., J.M.L., H.B.W. and J.J.Z. participated in the data analysis. E.G. contributed on discussion and writing of theoretical interpretation of the results.
Y.D.W., J.H., Y.C.T., P.W., H.T.C. and E.G. contributed to the writing of the manuscript.
All authors discussed the contents and form the final version of the paper.
\item[Competing Interests] The authors declare that they have no competing financial interests.
\end{addendum}
\newpage


\clearpage
\newpage

\begin{methods}
\setcounter{figure}{0}
\captionsetup[figure]{labelfont={bf},labelformat={default},labelsep=period,name={Extended Data Fig.}}

\subsection{Dataset and search procedures}

In the context of traditional methods, we conducted a refined search on the dataset obtained from 59.5 hours of FRB 20121102A observations in 2019. While previous analyses focused on brighter bursts, the current search shifts toward systematically identifying weaker signals at lower detection thresholds. We performed a single-burst search using Presto with a finer grid: the parameter space ranges from DM = 540 to 590 pc cm$^{-3}$, with a DM step of 0.02 pc cm$^{-3}$, compared to the parameter space used in Ref.~\cite{Li_2021}, which ranges from DM = 400 to 650 pc cm$^{-3}$, with a DM step of 0.2 pc cm$^{-3}$. Additionally, we reduced the SNR threshold from \( S_{\text{peak}}/\text{Noise} \geq 7 \) to 4. Based on the peak distribution of the SNR-DM plots for each candidate, we filtered out candidate bursts and manually inspected the dynamic spectra of each to ensure the removal of radio frequency interference (RFI) events. Furthermore, we increased the downsampling rate when plotting the dynamic spectra to maximize the identification of weak bursts in the dataset. Despite these refinements to the traditional procedure, the deep search was ultimately based on the AI algorithm EDEN we developed.

We propose the use of a neural network for end-to-end signal prediction. To enhance the model's predictive capability, we incorporate Teacher-Student Learning and Positive-Unlabeled (PU) Learning during the training process.

\subsubsection*{{i)} End-to-End FRB Detection through Image Recognition}

To enhance the efficiency and accuracy of Fast Radio Burst (FRB) detection, we propose an end-to-end approach that mitigates the biases and limitations inherent in traditional statistical methods. This novel strategy enables direct detection of FRBs, significantly reducing the reliance on pre-processed statistical data and the computational overhead associated with conventional methods.

To achieve this, we model the FRB detection problem as an image recognition task. As shown in Figure \ref{fig:ai_versus_heimdall}, the neural network used in this approach receives de-dispersed time-frequency data, which, when divided along the time dimension, resemble images. This conceptualization enables the application of advanced image recognition techniques to FRB detection. The training images, denoted as ${\mathbf{x}_i}, i=1\dots n$, along with the corresponding labels and arrival times, $\mathbf{y}_i$ and $\mathbf{t}_i$, are processed through the network, denoted as $\mathcal{N}$. The loss function that guides the network is formulated as:
\begin{equation}
	\mathcal{L} = \frac{1}{n}\sum_{i=1}^{n}\mathcal{H}_{ce}(\mathcal{N}(\mathbf{x}_i), \mathbf{y}_i)
\end{equation}
Here, $\mathcal{H}_{ce}(a,b) = -[a\log(b) + (1-a)\log(1-b)]$ is the cross-entropy loss function. By using this loss function, the network is trained to efficiently detect and predict the arrival times of FRB signals in the data.

This end-to-end approach, by eliminating the need for dedispersion and other pre-processing steps commonly used in traditional FRB detection methods such as PRESTO, significantly enhances detection efficiency. In PRESTO-based methods, the dedispersion step alone can consume up to 70\% of the total computational time. By directly analyzing the data, our method offers a more streamlined and potentially faster alternative for FRB detection.

\subsubsection*{{ii)} Implementing Teacher-Student Learning for FRB Detection}

Directly identifying Fast Radio Bursts (FRBs) from undedispersed images presents significant challenges, primarily due to the high noise levels in the raw data. This noise can obscure FRB signals, complicating the neural network's ability to differentiate between noise and genuine FRB signals. Additionally, the dispersion characteristic of FRBs is an intrinsic feature that must be incorporated into the learning process. To address these challenges, we propose a two-stage teacher-student scheme~\cite{hinton2015distilling}, which allows the network to effectively learn from undedispersed images.

In the first stage, we train a teacher network, $\mathcal{N}_T$, using dedispersed data, $\mathbf{x}_D$. This step is relatively straightforward, as the dedispersed data exhibit less noise, making them more conducive to learning FRB patterns. In the second stage, we train the student network, $\mathcal{N}_S$, using undedispersed data, $\mathbf{x}$. In this stage, the teacher network guides the student's learning process. Specifically, we input $\mathbf{x}$ into the student network, $\mathcal{N}_S$, and the corresponding dedispersed data, $\mathbf{x}_D$, into the teacher network, $\mathcal{N}_T$. This method allows for the transfer of the teacher network’s ability to detect signals in dedispersed data to the student model, thereby equipping the student model with the capacity to detect FRBs in undedispersed data:
\begin{equation}
	\label{Fcn7}
	\mathcal{L}_{KD} = \frac{1}{n}\sum\limits_{i} \mathcal{H}_{cross}(\mathbf{y}_S^i,\mathbf{y}_T^i).
\end{equation}
Here, $\mathcal{H}_{cross}$ is the cross-entropy loss, and $\mathbf{y}_T^i = \mathcal{N}_T(\mathbf{x}^i)$ and $\mathbf{y}_S^i = \mathcal{N}_S(\mathbf{x}_D^i)$ represent the outputs of the teacher network, $\mathcal{N}_T$, and the student network, $\mathcal{N}_S$, respectively. Through this knowledge transfer technique, the student network can be optimized without relying on the specific architecture of the given network.

By incorporating teacher-student interactions, the proposed method enables FRB detection without the need for dedispersion, significantly improving detection efficiency. In addition to efficiency, effectiveness is equally crucial. Our goal is to leverage deep learning to identify more samples than those detected by conventional methods. Therefore, the optimization problem extends beyond a simple binary classification task. The data that conventional algorithms fail to detect as FRBs are essentially unlabeled, comprising both positive and negative instances. This transforms the problem into one of learning from positive and unlabeled samples.

\subsubsection*{{iii)} Positive and Unlabeled Learning in FRB Detection}

The concept of positive and unlabeled (PU) learning~\cite{liu2003building} provides a practical solution to the challenges in FRB detection when only positive (FRB) and unlabeled (potentially FRB or non-FRB) data are available. The original PU algorithm is designed for binary classification problems and defines the expected risk \(R(\mathcal{N})\) for a classifier \(\mathcal{N}\) over inputs \(\mathbf{x}\) as:
\begin{equation}
	\begin{aligned}
		R(\mathcal{N}) &= \pi R_1(\mathcal{N}) + (1-\pi) R_{-1}(\mathcal{N}) \\
		&= \pi P_1(\mathcal{N}(\mathbf{x}) \neq 1) + (1-\pi) P_{-1}(\mathcal{N}(\mathbf{x}) \neq -1),
	\end{aligned}
\end{equation}
where \(\pi = P(\mathbf{y} = 1)\) represents the class prior for the positive class, and \(P_1\) and \(P_{-1}\) denote the marginal probabilities of positive and negative samples, respectively. Given the absence of labeled negative data, direct minimization of \(P_{-1}\) is infeasible. Instead, the approach focuses on indirectly minimizing \(P_{-1}\) by utilizing the probability of unlabeled samples \(P_U\). The expected risk on unlabeled data is thus formulated as:
\begin{equation}
	\begin{aligned}
		R_U(\mathcal{N}) = & P_U(\mathcal{N}(\mathbf{x}) = 1) \\
		= & \pi P_1(\mathcal{N}(\mathbf{x}) = 1) + (1-\pi) P_{-1}(\mathcal{N}(\mathbf{x}) = 1) \\
		= & \pi (1 - R_1(\mathcal{N})) + (1-\pi) R_{-1}(\mathcal{N}),
	\end{aligned}
\end{equation}
Consequently, the total risk \(R(\mathcal{N})\) can be rewritten as:
\begin{equation}
	\begin{aligned}
		R(\mathcal{N}) &= \pi R_1(\mathcal{N}) + (1-\pi) R_{-1}(\mathcal{N}) \\
		&= \pi R_1(\mathcal{N}) - \pi(1 - R_1(\mathcal{N})) + R_U(\mathcal{N}) \\
		&= 2\pi R_1(\mathcal{N}) + R_U(\mathcal{N}) - \pi.
	\end{aligned}
\end{equation}
By applying this approach, we effectively address the binary PU problem within the context of FRB detection. Specifically, this method is adapted for training the teacher network, thereby significantly enhancing the performance of the student network in detecting FRBs from undedispersed data.

\subsubsection*{{iv)} Experiments}

The data used for training and validation are detailed as follows: four data sets were collected from FAST, corresponding to FRB 20121102A, FRB 20180301A, FRB 20190520B, and FRB 20201124A. FRB 20121102A consists of 51 days of data collected from late August to October 2021. FRB 20180301A includes 2 days of data from March 2021. FRB 20190520B comprises 11 days of data spanning from April to December 2020. FRB 20201124A contains 4 days of data from late September 2021 and 5 days from February 2022. The collected data are two-dimensional (Frequency \& Time) and stored in `fits' format.

After collection, the data undergoes preprocessing on the `fits' files. First, the data is dedispersed to retain the original, non-dispersed form of the bursts. Next, we simulate the dispersion effect by applying a simulated Dispersion Measure (DM), which is uniformly sampled from the range [100, 1500]. Finally, the original `fits' data is segmented into clips of 2500 time units, with each clip overlapping the subsequent one by 1250 units. This overlap ensures that each burst is fully captured within at least one clip.

For training, we employ a deepened ResNet~\cite{RESNET_2016} architecture with 68 layers. The models are optimized using an SGD~\cite{SGD_1998} optimizer with a learning rate of 0.1, a batch size of 256, and a threshold of 0.5. Both the teacher and student models are trained for 200 epochs. The experiments are conducted on NVIDIA V100 GPUs using PyTorch~\cite{paszke2019pytorch}.

The application of EDEN to the FRB 20121102A dataset demonstrates its effectiveness in identifying signals. Figure \ref{fig:weak_signals} presents three representative examples of newly detected bursts: a weak and narrow burst, a wide burst, and one detected amidst strong radio frequency interference (RFI). These signals, distinguished by their unique morphology or the presence of significant environmental interference, are typically overlooked by traditional algorithms. The successful detection of such signals by EDEN highlights its superior detection capabilities.

\subsubsection*{{v)} Simulations}

To further highlight the superiority of EDEN and conduct a more comprehensive analysis, we utilize simulated data to evaluate the capabilities of our model. Typical mock bursts with varying morphologies are shown in Figure~\ref{fig-method-simu} (Panels (a) and (c)). The simulation enables a quantitative assessment of EDEN's performance under different signal conditions, offering valuable insights into its detection capabilities and potential limitations.

The simulated sample of 10,000 FRB bursts exhibits the following parameter distributions: The SNR distribution follows a normal distribution with mean \(\mu=6\) and standard deviation \(\sigma=2\). The bandwidth distribution is characterized by \(\mu=200\) MHz and \(\sigma=150\) MHz. The central frequencies show a normal distribution centered at 1250 MHz (\(\sigma=100\) MHz), with values truncated at the 1000-1500 MHz boundaries to reflect typical observational constraints.
The burst width distribution comprises three distinct sub-populations: (1) a narrow component (1/3 of sample) with \(\mu=8\) ms (\(\sigma=3\) ms), (2) an intermediate component (1/3) with \(\mu=70\) ms (\(\sigma=30\) ms), and (3) a broad component (1/3) showing \(\mu=400\) ms (\(\sigma=200\) ms). This tri-modal width distribution was designed to test the model's ability to detect bursts with extreme wide widths.

A radar chart (Figure \ref{fig:ai_versus_heimdall}) has been constructed using the normalized statistics from Table~\ref{tab:quant_comparison} to compare the performance of EDEN and Heimdall across seven dimensions. Overall, EDEN surpasses Heimdall in all aspects, with notable advantages in efficiency, precision, and comprehensive signal detection.

Efficiency reflects the model's speed, defined as the ratio of the total input time to the model’s runtime. EDEN operates 4.2 times faster than Heimdall, highlighting the advantages of the end-to-end approach.

Recall and Precision are critical metrics for assessing the detection capability of a model. Recall indicates the proportion of successfully detected signals out of all actual signals, while Precision measures the proportion of true signals among those identified by the model. The AI training set contains relatively few high SNR bursts, resulting in reduced sensitivity to high SNR bursts in the simulated data. To account for this bias, we selected simulated bursts matching the SNR distribution of the training set (SNR \(<\) 7) and computed their recall values separately. For this subset, EDEN achieved a recall of 0.730, compared to 0.509 for Heimdall. Although EDEN exhibits a slightly higher Recall than Heimdall, its Precision is 29 times greater, underscoring the significantly lower false positive rate of our model.

Additionally, we observed that EDEN is capable of detecting signals in specific domains where Heimdall fails. Notably, EDEN excels in identifying signals with low bandwidth (BW). For narrow signals with bandwidths below the median, EDEN detects 54\% more signals than Heimdall.

In terms of Time Width (TW), Heimdall’s performance is constrained by the coarse granularity of its detection for high-TW signals. This limitation stems from Heimdall’s traditional grid search approach, which uses $2^n$ filtering and has a fixed grid length. As a result, high-TW signals represent a weakness for Heimdall. In contrast, EDEN is not subject to these limitations.

To compare the burst width detection granularity between EDEN and Heimdall, we computed the reciprocal of the root mean square (RMS) of recall fluctuations. The reciprocal for EDEN and Heimdall are 0.346 and 0.233, indicating that the recall of EDEN is less influenced by variations in TW, demonstrating greater consistency across different TW ranges. Furthermore, EDEN shows a significant advantage over Heimdall in detecting high-TW signals. Traditional algorithms, such as Heimdall, rely on adaptive filtering with templates based on powers of 2, which exhibit higher detection efficiency for bursts with widths close to the template values. However, real burst widths vary continuously, posing challenges for matching bursts with widths between adjacent exponents. Heimdall uses templates with lengths of \(2^n \times 98.304\, \mu s\) to match burst widths, so we analyzed detection performance as a function of exponent \(n\) by calculating the EDEN-to-Heimdall detection ratio. The results show that EDEN outperforms Heimdall in most width bins, with a ratio greater than 1. Notably, the ratio increases significantly when \(n>12\), highlighting our model's finer granularity and its superior capability in detecting wider bursts.

We also evaluated the model’s performance in detecting extremely low signal-to-noise ratio (SNR) bursts. The results show that our EDEN achieves a recall rate of 35.3\% for bursts with SNR \(<\) 3, whereas Heimdall only reaches 5.9\%. This highlights EDEN’s superior capability in detecting low-SNR bursts, enabling a more complete and unbiased FRB detection.

\subsection{Burst comprehensive analysis}

\subsubsection*{{i)} Burst energy analysis}
Based on the proposed AI algorithm EDEN, we performed an extensive search on the dataset reported in Ref. \cite{Li_2021} for FRB 20121102A in 2019, detecting a total of 5,927 independent bursts, tripling the burst set and doubling the total isotropic equivalent energy. This substantial increase in the detected bursts provides a more complete picture of the total emitted energy, enabling a more stringent constraint on the magnetar energy budget. 

To calculate the total isotropic equivalent energy emitted during the 59.5 hours of observation over 47 days, we adopt Equation (9) in Ref.~\cite{Zhang_2018} to maintain consistency with Ref.~\cite{Li_2021}:
\begin{equation}
    E_{iso,c} = \sum (10^{39}\text{erg})\frac{4\pi}{1+z}(\frac{D_L}{10^{28}\text{cm}})^2(\frac{F}{\text{Jy}\cdot\text{ms}})(\frac{\nu_c}{\text{GHz}})
\end{equation}
,where \(F_{\nu} = S_{\nu} \times W_{\text{eff}}\) is the fluence and \(\nu_c\) is the central frequency of 1.25 GHz. \(S_{\nu}\) is the peak flux density, calibrated based on the baseline noise level and then measured upon this baseline, \(W_{\text{eff}}\) is the effective burst width, calculated by dividing the equivalent rectangle area of the burst by the maximum value of the spectrum after baseline subtraction. The luminosity distance \(D_L = 949\) Mpc corresponds to a redshift \(z = 0.193\) for FRB 20121102A \cite{Tendulkar_2017}. This yields a total energy of \(5.94 \times 10^{41}\) erg, nearly doubling the total energy of \(3.41 \times 10^{41}\) erg reported in Ref.~\cite{Li_2021}. 

When calculating the equivalent magnetar energy, it is important to note that the isotropic energy of FRBs is derived from the burst spectra~\cite{ZhangB_2023}. Since repeating bursts typically exhibit narrow-banded spectra, we use the observation bandwidth of \(\Delta\nu = 500\) MHz instead of the central frequency \(\nu_c\). Assuming a typical radiative efficiency of \(\eta_r \sim 10^{-4}\) and a total magnetic energy of a typical magnetar with \(B = 10^{15}\) G and \(R = 10\) km, given by \(E_B = B^2 R^3 / 6 = 1.7 \times 10^{47}\) erg, we obtain a total \(E_{iso,w}\) equivalent to 26.2\% of the available magnetar energy. A comparison with three other repeating FRBs—FRB 20190520B~\cite{Niu_2022}, FRB 20201124A~\cite{Xu_2022,ZhangYK_2022}, and FRB 20220912A~\cite{ZhangYK_2023}—based on FAST observations is shown in Figure~\ref{fig-main-magnetar}. Panel (a) compares the fraction of magnetar energy corresponding to the total isotropic energy of the four repeating FRB sources, with the stacked bar distinguishing the contributions from two observation periods for FRB 20201124A. Notably, the equivalent magnetar energy estimated in this study exceeds that of other sources, even when combining the two observations of FRB 20201124A. This rules out models with low radio radiative efficiency and provides the most stringent constraints on the energy origin model of magnetars. Panel (c) quantifies the discrepancies in the cumulative energy distributions shown in panel (b) by displaying the residual of the cumulative distribution and the residual per energy bin. The primary growth in energy occurs between \(10^{37}\) erg and \(10^{38}\) erg. Our new detections, consisting predominantly of faint bursts, significantly increase the total energy through their cumulative effect.

To consistently classify multi-peak bursts, we adopted the following criteria, in alignment with previous detection methods: two distinct bursts are considered separate if the drop between peaks and valleys in the profile exceeds 3\(\sigma\). However, a new criterion, stating that "the drop between peaks and valleys must exceed 5\(\sigma\), with adjacent burst intervals greater than the width of the preceding burst," appears to be more appropriate and effective.

We examined the bimodal burst rate-energy distribution on the new burst set. As shown in Figure \ref{fig-method-E-br}, the derived energies span four orders of magnitude, from \(10^{36}\) erg to \(10^{40}\) erg. The burst energy distribution remains bimodal, described by a combination of a log-normal function and a Cauchy-Lorentz function. The characteristic energy is \( 4.84 \times 10^{37}\) erg, which is in close agreement with the previous value of \( 4.8 \times 10^{37}\) erg. The fitted function is expressed as follows:
\begin{equation}
    N(E) = \frac{N_0}{\sqrt{2 \pi} \sigma_E E} \exp\left(-\frac{(\log E - \log E_0)^2}{2 \sigma_E^2}\right) + \frac{\epsilon}{1 + \left(\frac{E}{E_1}\right)^{\alpha_E}}
\end{equation}

The enlarged sample size did not alter the bimodal distribution in the time domain. The distribution consists of a peak centered around \(4.84 \times 10^{37}\) erg and a high-energy tail with a slope of -1.26. To test the robustness of the bimodal distribution fit, we also performed curve fitting using a single power law, a Cauchy function, and a log-normal function, respectively. The fitting parameters are listed in Table \ref{tab:parameters}. The \(R^2\) value for the bimodal function fit is 0.996, surpassing that of all other functions, indicating that the data cannot be adequately described by a single component. 

We also calculated the PCC for the two burst sets, yielding \(r(32) = 0.89, p = 1.33 \times 10^{-12}\). This indicates a highly similar morphological pattern between the two distributions, demonstrating a strong correlation between the two bimodal burst energy distributions.

\subsubsection*{{ii)} Bandwidth distribution}
Figure \ref{fig-method-bw} displays the distribution of burst bandwidths, along with a statistical histogram fitted to a log-normal function. The upper panel shows the bandwidth distributions of the two burst sets, represented in grey and red, respectively. Both distributions are fitted using a log-normal function, with dashed vertical lines indicating the peak values. The new burst set peaks at 77 MHz, suggesting the detection of narrower bursts. The newly detected bursts are primarily concentrated in the 50-120 MHz range, with only 2\% of bursts falling within the broader 300-500 MHz bandwidth, compared to 14\% for the previously detected bursts. The lower panel plots the individual burst bandwidths against their burst IDs, illustrating the temporal spread of bandwidths for both burst sets.

This plot illustrates the temporal variation of bandwidths across the entire sample, highlighting differences in the temporal distribution of newly detected bursts. The AI model employed in the new detection primarily identifies narrow-bandwidth bursts. In contrast, the traditional search pipeline must integrate across the entire bandwidth, which introduces multiple instances of the noise background for narrow-bandwidth bursts, thereby significantly compromising detection performance.

\subsubsection*{{iii)} Energy and waiting time distributions}

Figure \ref{fig-method-Eratio-wt} illustrates the energy ratio \( E_{i+1}/E_i \) and waiting time distribution. The energy ratio is defined as the energy of a subsequent burst divided by that of the previous burst. The waiting time for each burst is calculated within each \( E_{i+1}/E_i \) bin. The upper panel shows the distribution across all energy bands, while the lower panels present results for three specific energy ranges: \(E \leq 4 \times 10^{37}\) erg, \(4 \times 10^{37} \) erg \(< E < 3 \times 10^{38}\) erg, and \(E \geq 3 \times 10^{38}\) erg. As energy increases, the waiting time distribution exhibits a clear evolutionary trend.

The bimodal energy distribution can be qualitatively interpreted in terms of energy release in a magnetar through glitch-like events. Seismic energy injected into the magnetosphere may produce short-duration FRBs \cite{Lu2020}. A similarity exists between the spin glitch size $\Delta\nu_{\rm glitch}$ distribution of magnetars (see Figure 2 in Ref. \cite{Fuentes2017}) and the burst energy distribution of FRB 20121102A (see Figure \ref{fig-method-E-br}). Magnetic field evolution and a large spin-down rate, which assist in accumulating stresses \cite{Thompson1996}, along with superfluid interior dynamics \cite{Link1996}, lead to spin glitches that release energy into the magnetosphere via mechanical dissipation from crustal breaking and rotational dissipation from superfluid vortex creep heating, respectively.
The mechanical energy released during a crustquake is given by \cite{Baym1971}
\begin{equation}
    \Delta E_{\rm quake}=2\theta_{\rm cr}B\frac{\Delta\Omega}{\Omega}\cong10^{37}-10^{41} \mbox{erg},
\end{equation}
where $\theta_{\rm cr}\cong0.01-0.1$ \cite{Horowitz2009,Baiko2018} is the critical strain angle for crustal failure, $B\sim10^{48}$ erg \cite{Cutler2003,Zdunik2008} is a constant related to lattice rigidity, and $\Delta\Omega/\Omega\approx10^{-9}-10^{-6}$ \cite{Zhou2022} is the fractional amplitude of the rotational glitch. Recent calculations show that the maximum strain energy available in the neutron star crust due to rotation is $E_{\rm strain}\sim0.6\times10^{46}$\,erg \cite{Rencoret2021}. In reality, the intense magnetic field of a magnetar causes some fraction of this strain energy to be lost as plastic flow \cite{Thompson2017,Gourgouliatos2021}. Thus, an efficiency of $\eta=10^{-6}$ for converting crustal strain energy into seismic energy during a quake can account for the energy budget of the bursts observed from FRB 20121102A, with their concentration between $10^{37}-10^{41}$ erg.
The standard scenario of pulsar glitches \cite{Alpar1984} involves the collective discharge of superfluid vortex lines by unpinning from the lattice nuclei. During this process, part of the star's rotational energy is converted into heat, as described by \cite{erbil2019}
\begin{equation}
    \Delta E_{\rm sf}=I_{\rm sf}\omega_{\infty}\delta\Omega_{\rm sf},
    \label{sfenergy}
\end{equation}
where $I_{\rm sf}~10^{-2}I=10^{43}$\,ergs s$^{-2}$ is the moment of inertia of the crustal superfluid region, which drives a glitch spin-up event through the large-scale, collective unpinning of superfluid vortex lines, $\omega_{\infty}=\Omega_{\rm s}-\Omega_{\rm c}$ is the steady-state angular velocity lag between the angular velocities of the superfluid ($\Omega_{\rm s}$) and crust ($\Omega_{\rm c}$) components, and $\delta\Omega_{\rm sf}$ is the change in the superfluid's angular velocity due to vortex line unpinning. For the superfluid in the inner crust and outer core, the steady-state lag values are 0.01 rad s$^{-1}$ and 0.1 rad s$^{-1}$, respectively \cite{erbil2016}. The change in the superfluid angular velocity during a glitch is related to the number of vortices unpinned, $N_{\rm V}$, by $\delta\Omega_{\rm sf}=(N_{\rm V}\kappa)/(2\pi R^{2})$, where $\kappa=2\times10^{-3}$\,cm$^{2}$s$^{-1}$ is the vorticity quantum associated with each line, and $R\approx10$\,km is the neutron star radius. Post-glitch timing fits \cite{erbil2022} show that the number of vortices discharged during glitches is remarkably constant, $N_{\rm V}=(1-5)\times10^{13}$, yielding $\delta\Omega_{\rm sf}\lesssim10^{-2}$\,rad s$^{-1}$. Thus, Equation (\ref{sfenergy}) gives $\Delta E_{\rm sf}=3.2\times10^{38}-1.6\times10^{41}$\,erg. Therefore, the bimodal energy distribution can be explained by the simultaneous presence of crustquake and superfluid-induced energy release mechanisms to the magnetosphere in a young magnetar. If this interpretation is correct, future bursts with energies $\gtrsim10^{41}$\,erg may be detected in the FRB20121102A source. The clear evolutionary trend of the energy ratio \( E_{i+1}/E_i \) clustering around 1 as burst energy increases can be understood in terms of the spin glitches observed in normal radio pulsars. Large pulsar glitches are occasionally preceded or followed by smaller events \cite{Espinoza2021}, with $E_{i+1}/E_{i}>0$ and $E_{i+1}/E_{i}<0$ corresponding to precursors and post-shocks, similar to those seen in earthquakes. As burst energy increases, the large scatter in $E_{i+1}/E_{i}$ diminishes, and the ratio converges to 1, as there exists a physical upper limit to crustal strain and the superfluid reservoir.

The bimodal waiting time distribution, peaking at tens of milliseconds and several seconds, is also evident for FRB 20201124A \cite{ZhangYK_2022} and FRB 20220912A \cite{Konijn2024}, suggesting that the same two physical mechanisms are responsible for the two extremes in the observed distribution across different sources. Generally, events occurring on timescales shorter than the light-crossing time of the light cylinder radius, $R_{\rm LC}$, i.e., $\tau\lesssim R_{\rm LC}/c$, are associated with conditions prevailing in the magnetosphere, while longer timescales, $\tau\gtrsim R_{\rm LC}/c$, are linked to processes related to the neutron star's interior dynamics. A hidden large toroidal magnetic field within the interior would deform the magnetar's crust from a relaxed spherical shape. Magnetar precession has been proposed as a plausible mechanism for repeating FRBs \cite{Levin2020,Li2021,Wasserman2022}. Precession may trigger frequent crustquakes on a timescale \cite{Barsukov2010}
\begin{equation}
    t_{\rm precession}=\frac{P}{\cos{\chi}\epsilon_{\rm B}},
    \label{pretime}
\end{equation} 
where $P$ is the magnetar's spin period, $\chi$ is the wobble angle, and $\epsilon_{\rm B}$ is the ellipticity of the magnetar's crustal distortion due to the large internal magnetic field. The inferred wobble angles for PSR B1828-11 \cite{Cutler2003} and the transient radio-emitting magnetar XTE J1810-197 \cite{Desvignes2024} are small, approximately $\sim3^{\circ}$ and $\sim20^{\circ}$, respectively. Therefore, we take $\cos{\chi}\approx1$ in our estimations. The ellipticity resulting from the magnetic distortion of the star is given by \cite{Akgün2008}
\begin{equation}
  \epsilon_{\rm B}=\xi\frac{B_{\rm in}^{2}R^{4}}{GM^{2}},
\end{equation}
where $B_{\rm in}$ is the crustal magnetic field, $R\approx10$\,km is the neutron star radius, $G$ is the gravitational constant, $M=1.4M_{\odot}$ is the neutron star mass, and $\xi=10-100$ is a constant determining the degree to which the internal magnetic field deviates from a simple dipolar geometry. For these fiducial values, $\epsilon_{\rm B}\cong2\times10^{-5}(B_{\rm in}/10^{15}\mbox{G})$, and Equation (\ref{pretime}) gives $t_{\rm precession}=5\times10^3(P/0.1\mbox{s})$\,s. Under certain conditions, precession can mediate the unpinning of vortex lines in parts of the crustal superfluid \cite{Link2002}. For superfluid-mediated glitches, the repetition timescale is given by \cite{Alpar1984,erbil2020}
\begin{equation}
    t_{\rm sf,waiting\,time}=\frac{\delta\Omega_{\rm s}}{|\dot{\Omega}|}=(500-2500)\left(\frac{P}{0.1\mbox{s}}\right)^{2}\left(\frac{\dot{P}}{10^{-8}}\right)^{-1}\mbox{s},
\end{equation}
where $|\dot{\Omega}|=2\pi\dot{P}/P^{2}$ is the absolute magnitude of the spin-down rate, with $P$ and $\dot{P}$ being the spin period and its first derivative of the underlying magnetar. In the crustquake scenario, the timescale between successive events is expressed in terms of the magnitude of the preceding quake \cite{Baym1971}
\begin{equation}
  t_{\rm quake}=\frac{A^{2}\tau_{\rm sd}}{B I_{0}\Omega^{2}}\frac{\Delta\Omega}{\Omega}\gtrsim6\times10^{6} \mbox{s},
\end{equation}
where $A\sim GM^{2}/R\sim10^{53}$\,erg is a constant of the order of the gravitational binding energy of the neutron star, $\tau_{\rm sd}=P/(2\dot{P})$ is the characteristic (spin-down) timescale, $I_{0}\sim 10^{45}$\,ergs s$^{-2}$ is the moment of inertia of the non-rotating spherical star, and $\Omega=2\pi/P$ is the angular velocity of the crust's rotation. Twisting of the magnetic field lines within the magnetar \cite{Thompson2000} and the pinning of vortex lines to magnetic flux tubes \cite{Ruderman1998} are two mechanisms that shorten the repetition time between successive quakes by adding magnetic stresses constructively to the crustal strain due to magnetar spin-down. Thus, free precession, superfluid-induced glitches, and crustquakes driven by spin slowdown and magnetic field evolution may account for the observed waiting times ranging from $10$\,s to $10^{6}$\,s in magnetars. For millisecond-scale waiting times, magnetospheric mechanisms responsible for giant bursts in canonical neutron stars and millisecond pulsars are likely dominant. Nonlinear plasma processes in the magnetosphere grow rapidly, discharge quickly, and can repeat efficiently. The propagation of drift waves at the magnetospheric boundary, which transforms accumulated energy into narrow-band emissions \cite{Machabeli2019}, and the reconnection of plasmoids beyond the light cylinder \cite{Cerutti2013}, can produce energetic bursts with very short timescales, potentially explaining the clustering of waiting times at the millisecond extreme of the distribution shown in Figure \ref{fig-method-Eratio-wt}.

\subsubsection*{{iv)} DM variation}
We used the \textit{DM-Power} algorithm (https://github.com/hsiuhsil/DM-power), as described by Ref. \cite{Lin_2023dmpower}, for dedispersion and precise determination of the DM value. Figure \ref{fig-method-dm} displays the DM distribution for each burst, showing that values are concentrated between 560 and 567 pc cm$^{-3}$. The mean DM prior to MJD 58760 is 563.4 pc cm$^{-3}$ with a variance of 0.04 pc cm$^{-3}$, while after this date, the mean DM increased by only 0.05 pc cm$^{-3}$. However, the reduced number of bursts after MJD 58760 resulted in a significantly larger variance of 0.4 pc cm$^{-3}$, which is 10 times greater than the variance before MJD 58760. The right panel shows histograms of the newly detected bursts (in blue) and previous bursts (in grey). The new detections are largely consistent with the prior results, with the exception of an increase in their number.

\subsubsection*{{v)} Correlation analysis}
The pairwise scatter plot matrix in Figure \ref{fig-method-corr} illustrates the correlations between the dispersion measure (DM), the logarithm of \(W_{\rm eff}\) (lg \(W_{\rm eff}\)), the logarithm of flux (lg Flux), the logarithm of energy (lg E), and bandwidth.

The scatter plots involving DM show no significant trend with the other variables, with a clustered pattern suggesting a weak correlation between DM and these variables. In contrast, the plot of \(\log(\text{Flux})\) versus \(\log(E)\) reveals a strong positive correlation, indicating that higher fluxes are associated with higher energies. A positive correlation is also observed between energy and \(\log(W_{\text{eff}})\), though this relationship is weaker compared to the one with flux. The flux range spans approximately 3.5 orders of magnitude, whereas \(\log(W_{\text{eff}})\) spans only 1.5 orders of magnitude. Therefore, flux primarily governs the energy distribution, in contrast to \(\log(W_{\text{eff}})\), which is similar to FRB 20201124A and FRB 20240114A, but differs from FRB 20220912A, where \(\log(W_{\text{eff}})\) dominates the energy distribution.

Pearson correlation analysis reveals a moderate positive correlation between bandwidth and both \(\log \mathrm{Flux}\) (\(r = 0.491\), \(p < 0.0001\)) and \(\log \mathrm{E}\) (\(r = 0.443\), \(p < 0.0001\)). These results indicate that as bandwidth increases, both \(\log \mathrm{Flux}\) and \(\log \mathrm{E}\) also increase, suggesting that higher energies are associated with larger bandwidths. The extremely low p-values for both correlations indicate that these relationships are statistically significant and not attributable to random chance. bursts within the 400-500 MHz bandwidth range predominantly have energies between \(10^{38}\) erg and \(10^{40}\) erg. The newly detected, weaker bursts primarily exhibit narrower bandwidths, thereby populating the parameter space at lower thresholds. Furthermore, the absence of a linear relationship between flux and \(W_{\text{eff}}\) does not support the asteroid collision model proposed by Ref. \cite{Dai_2020}.

\newpage

\begin{figure*}
    \centering
    \includegraphics[scale=0.5]{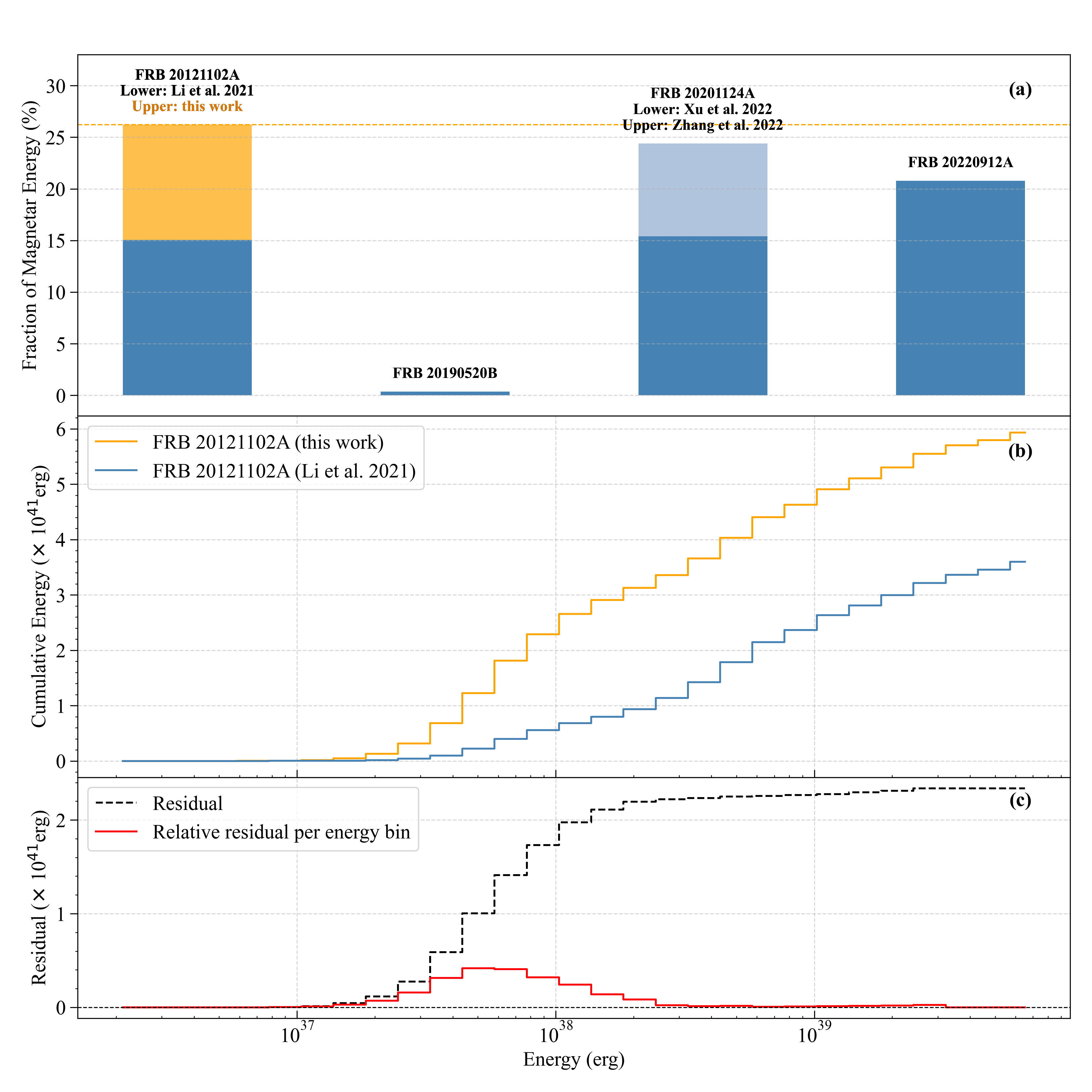}
    \caption{ \textbf{Analysis of the corresponding fraction of magnetar energy and total energy of the bursts.} Panel (a): The fraction of magnetar energy corresponding to the total energy of various repeating FRB sources observed by FAST. Four sources are included, represented by blue bars (for data from the literature) and orange bars (for this work). The stacked bars for FRB 20201124A highlight the combined contributions from two observation periods. The total isotropic burst energy derived in this study accounts for 26\% of the available energy of a typical magnetar, surpassing the corresponding energy obtained from other sources (including the sum of the two observation periods of FRB 20201124A). The values used in panel (a) are as follows: 26.2\%, 0.4\%, 24.4\%, 20.8\%. Panel (b): Cumulative energy distribution for two burst sets of FRB 20121102A. Panel (c): The distribution of residuals between the cumulative energy curves shown in panel (b). The black dashed line represents the cumulative residuals, while the red solid line represents the relative residuals per energy bin.}
    \label{fig-main-magnetar}
\end{figure*}

\begin{table}
    \centering
    \begin{tabularx}{1\textwidth}{cccc}
        \hline\hline
        Function & Fitting parameter & Energy range(erg) & \(\bar{R}^{2, a}\) \\
        \hline
        Power law &
        \begin{tabular}{c}
            \(\gamma= -0.07\pm 0.11\)  \\
            \(\gamma= -1.44\pm 0.12\)
        \end{tabular} &
        \begin{tabular}{c}
            \(4\times 10^{36} \leq E \leq 8\times 10^{39}\)   \\
            \(1\times 10^{38} \leq E \leq 8\times 10^{39}\)   
        \end{tabular}
        & \begin{tabular}{c}
            -0.041  \\
            0.956
         \end{tabular} \\
        \hline
        Lognormal & 
        \begin{tabular}{c}
            \(E_0=6.63\times 10^{37}\)(erg) \\
            \(\sigma_E=0.58\) \\
            \(N_0=6.03\times 10^{38}\)
        \end{tabular} & 
        \(4\times 10^{36} \leq E \leq 8\times 10^{39}\) & 0.985 \\
        \hline
        Cauchy & 
        \begin{tabular}{c}
            \(E_0=4.95\times 10^{38}\)(erg) \\
            \(\alpha_E=3.36 \pm14.77\)
        \end{tabular} & 
        \(4\times 10^{36} \leq E \leq 8\times 10^{39}\) & -0.045 \\
        \hline
        Lognormal+Cauchy &
        \begin{tabular}{c}
           \(E_0=5.87\times 10^{37}\)(erg)  \\
           \(\sigma_E=0.54\) \\
           \(N_0 = 4.81\times 10^{38}\) \\
           \( E_1 = 3.71\times 10^{38} \)(erg) \\
           \( \alpha_E=1.26 \pm 0.3\)
        \end{tabular} &
        \(4\times 10^{36} \leq E \leq 8\times 10^{39}\) & 0.996 \\
        \hline
    \end{tabularx}
    \caption{The fitted parameters.}
    \label{tab:parameters}
\begin{tablenotes}
\item[a] $^a$Adjusted coefficient of determination. \(\bar{R}^2=1-\frac{SS_{res}}{SS_{tot}}(\frac{n-1}{n-p-1})\), where \(SS_{res}\) is the residual sum of squares and \(SS_{tot}\) is the total sum of squares which is proportional to the data variance.
\end{tablenotes}
\end{table}

\begin{figure*}
	\centering
	\includegraphics[width=1\linewidth]{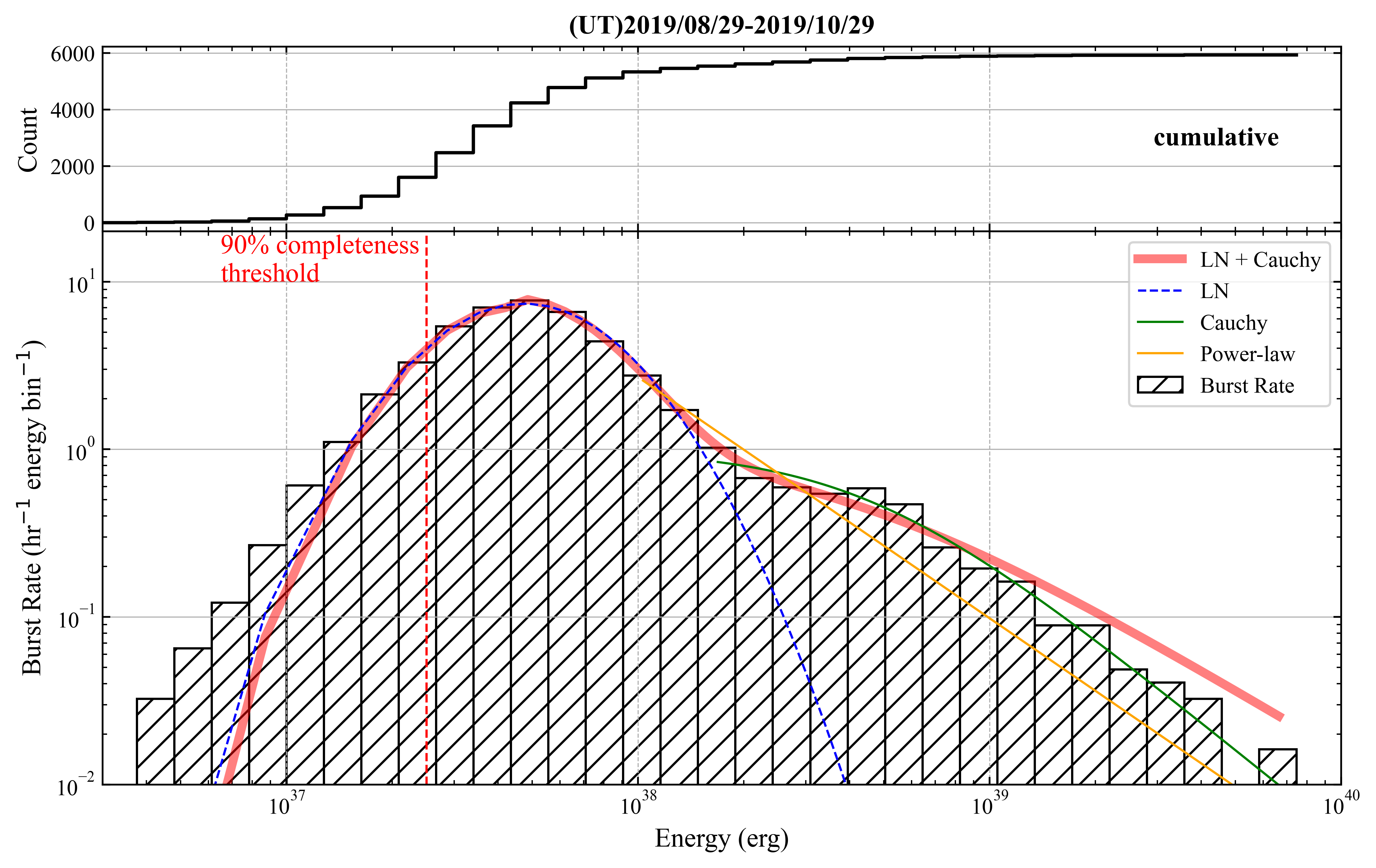}
	\caption{\textbf{Burst rate distribution of energy for FRB 20121102A bursts detected in a deep search.} The histogram illustrates the burst rate distribution, with the dashed blue line representing the Log-Normal fit, the solid green line indicating the Cauchy fit, and the solid yellow line showing the power-law fit in the energy range \(1\times 10^{38} \leq E \leq 8\times 10^{39}\). The solid red line denotes the combined Log-Normal and Cauchy fit. The vertical dashed red line marks the 90\% completeness threshold. The upper subplot presents the cumulative count of burst events over the specified time period.}
	\label{fig-method-E-br}
\end{figure*}

\begin{figure*}
	\centering
	\includegraphics[width=1\linewidth]{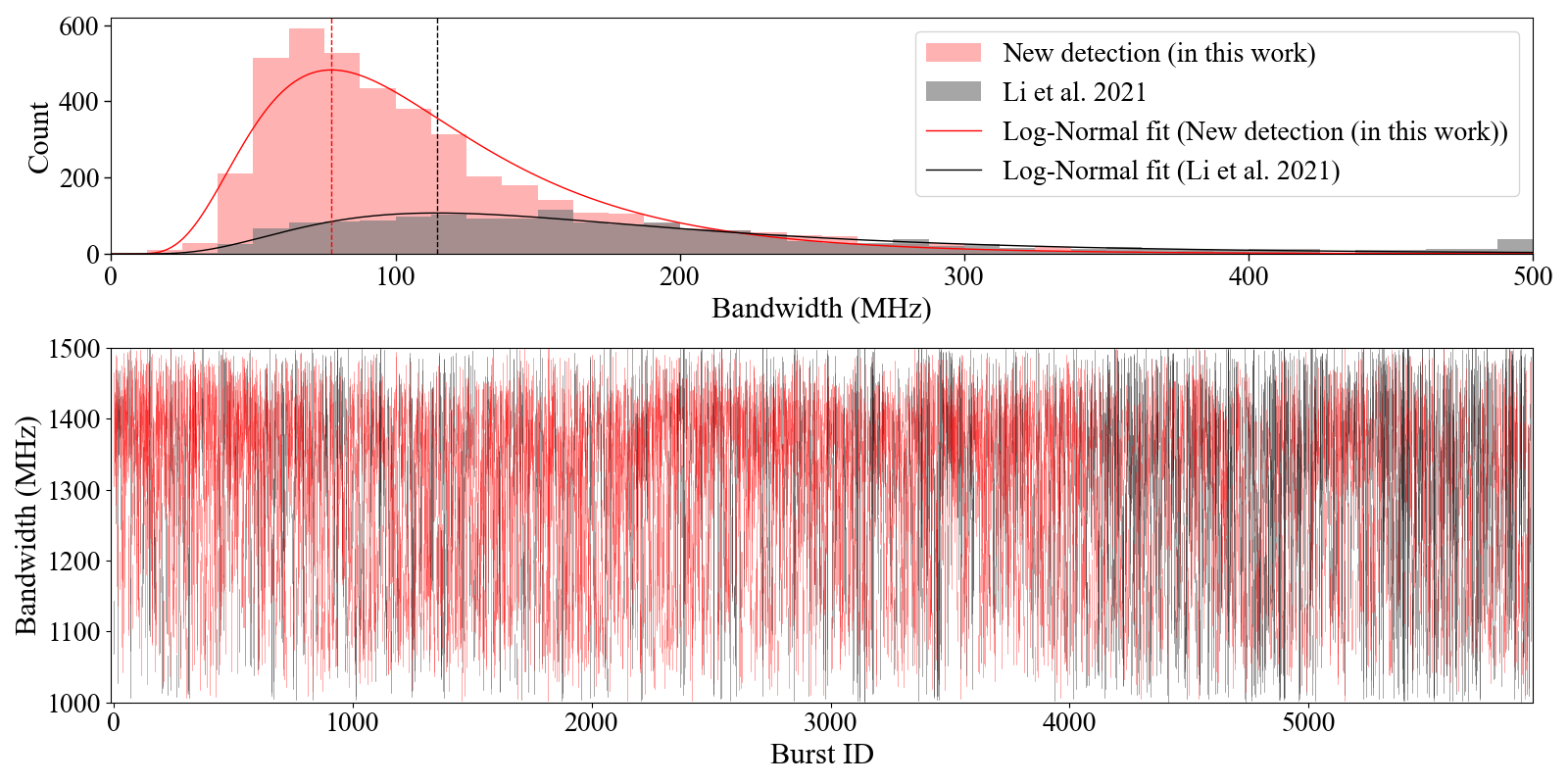}
	\caption{\textbf{Bandwidth distribution of the bursts.} The upper panel displays the histogram of bandwidths (in MHz) along with log-normal fits for two different burst sets: the newly detected bursts, shown in red, are fitted with a red log-normal curve, while the bursts reported by Ref.\cite{Li_2021} are depicted in gray with the corresponding black log-normal fit. The lower panel illustrates the bandwidth distribution for all bursts, with newly detected bursts represented by red lines and previous detections by gray lines. It is apparent that the new algorithm identifies bursts with generally narrower bandwidths compared to the previous search.}
	\label{fig-method-bw}
\end{figure*}

\begin{figure*}
	\centering
	\includegraphics[width=1\linewidth]{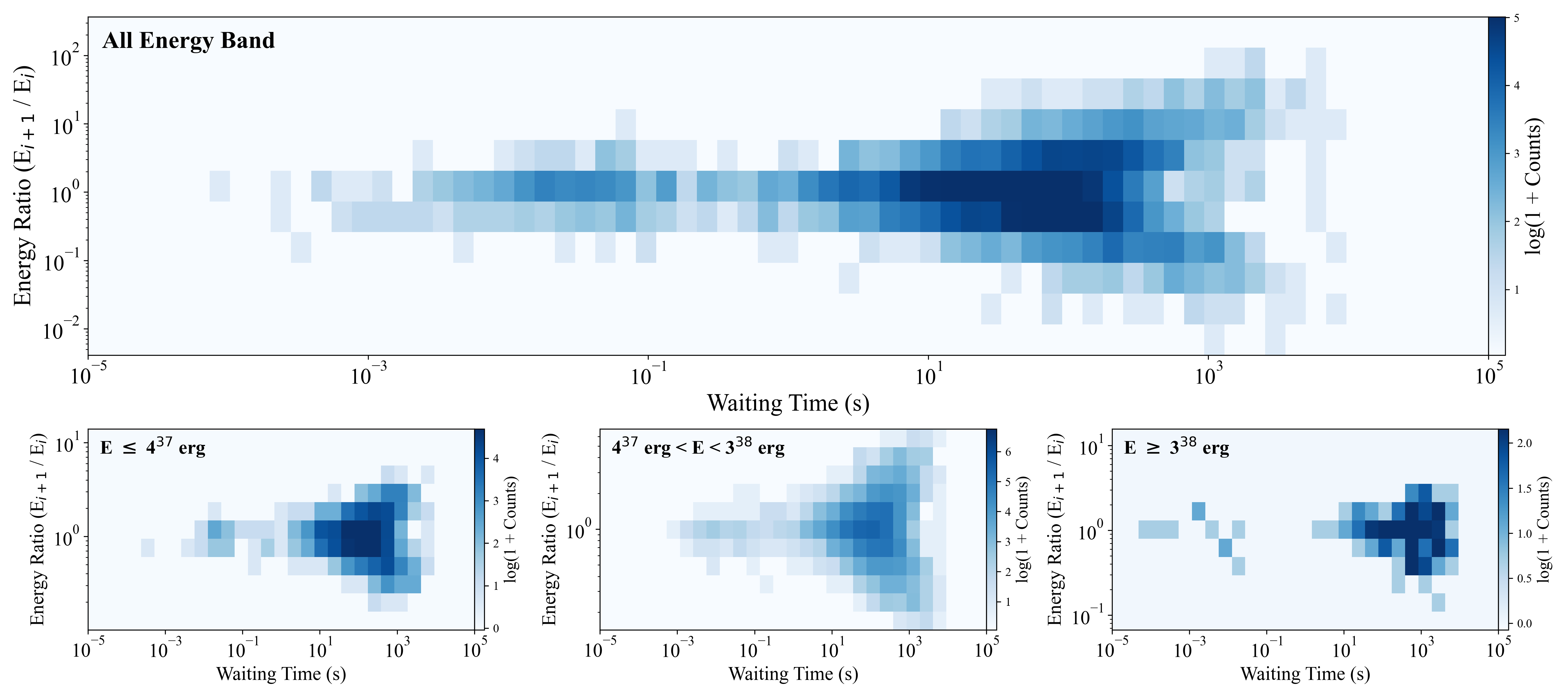}
	\caption{\textbf{Energy ratio \( \mathbf{E_{i+1}/E_i} \) and waiting time distribution of the bursts.} The energy ratio is defined as the energy of a subsequent burst divided by that of the preceding burst. The energy ratio bins are then categorized according to their respective values. The waiting time for bursts is calculated within each bin. The upper panel shows the distribution across all energy bands, while the three lower panels, from left to right, display the results for three specific energy ranges: \(E \leq 4 \times 10^{37}\) erg, \(4 \times 10^{37}\) erg \( < E < 3 \times 10^{38}\) erg, and \(E \geq 3 \times 10^{38}\) erg. As energy increases, the waiting time distribution clearly exhibits an evolutionary trend.}
	\label{fig-method-Eratio-wt}
\end{figure*}

\begin{figure*}
	\centering
	\includegraphics[width=1\linewidth]{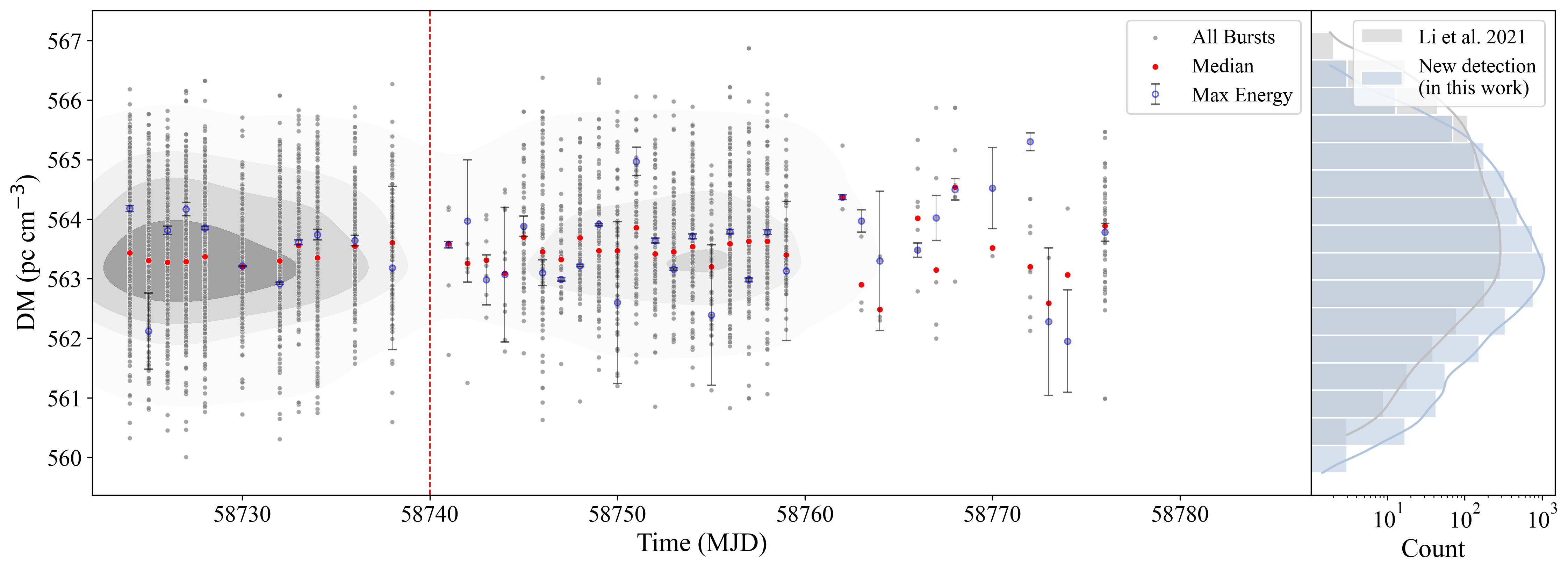}
	\caption{\textbf{Dispersion Measure (DM) distribution of the bursts.} The gray dots represent the DM values for bursts detected at each epoch. The red and blue dots indicate the median DM and the DM of the highest energy burst for each epoch, respectively. The DM distribution shows no significant variation over the course of the observation. The right panel displays histograms of burst distributions for both the new and previously detected bursts, revealing negligible differences between the two burst sets.}
	\label{fig-method-dm}
\end{figure*}

\begin{figure*}
	\centering
	\includegraphics[width=1\linewidth]{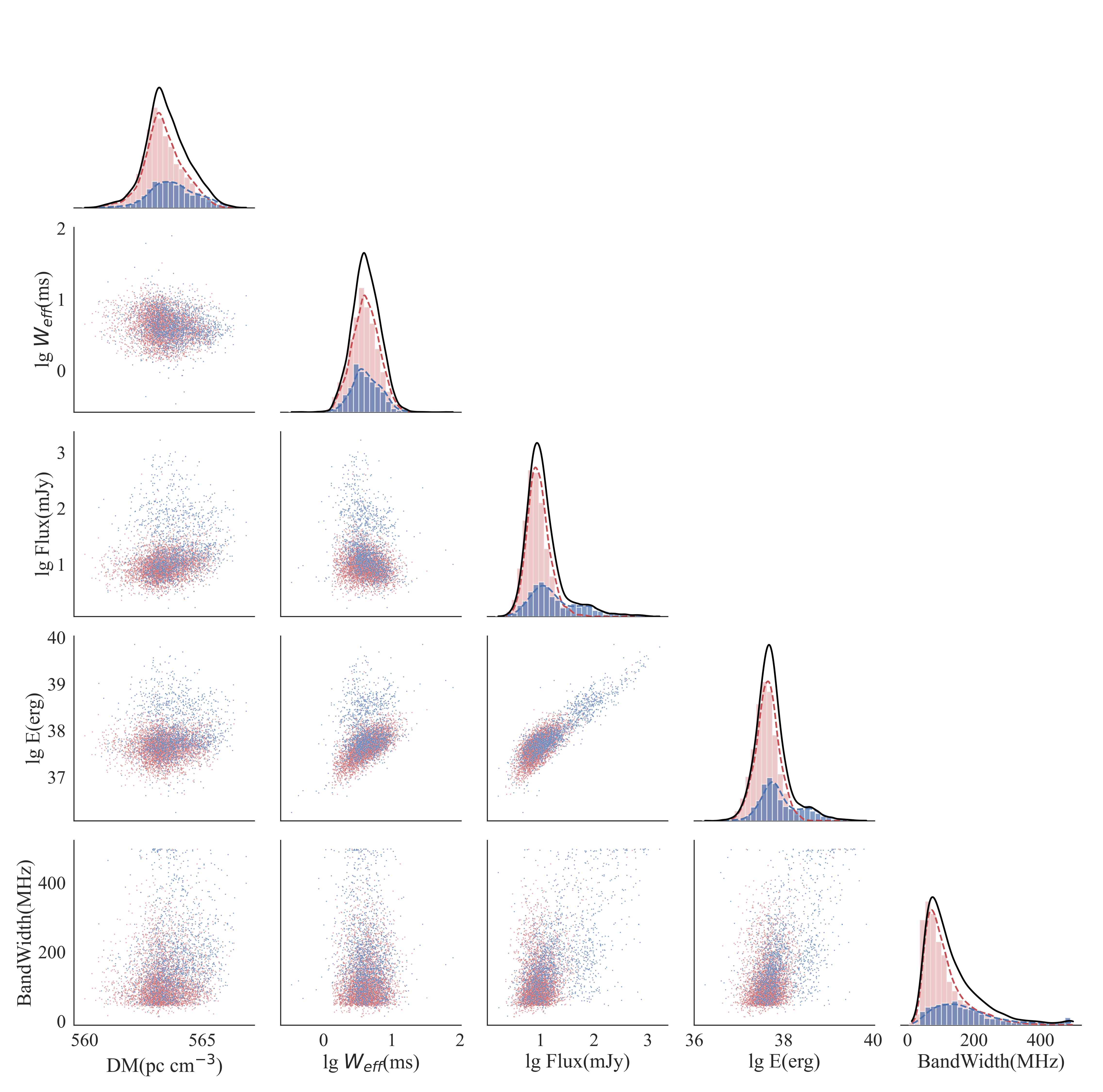}
	\caption{\textbf{Correlation matrix of five variables.} The pairwise scatter plot matrix illustrates the correlations between dispersion measure (DM), logarithm of \(W_{\text{eff}}\) (lg \(W_{\text{eff}}\)), logarithm of flux (lg Flux), logarithm of energy (lg E), and bandwidth. Pearson correlation analysis reveals a moderate positive correlation between bandwidth and both $\lg \mathrm{Flux}$ ($r = 0.491$, $p < 0.0001$) and $\lg \mathrm{E}$ ($r = 0.443$, $p < 0.0001$).}
	\label{fig-method-corr}
\end{figure*}

\begin{figure*}
	\centering
	\includegraphics[width=1\linewidth]{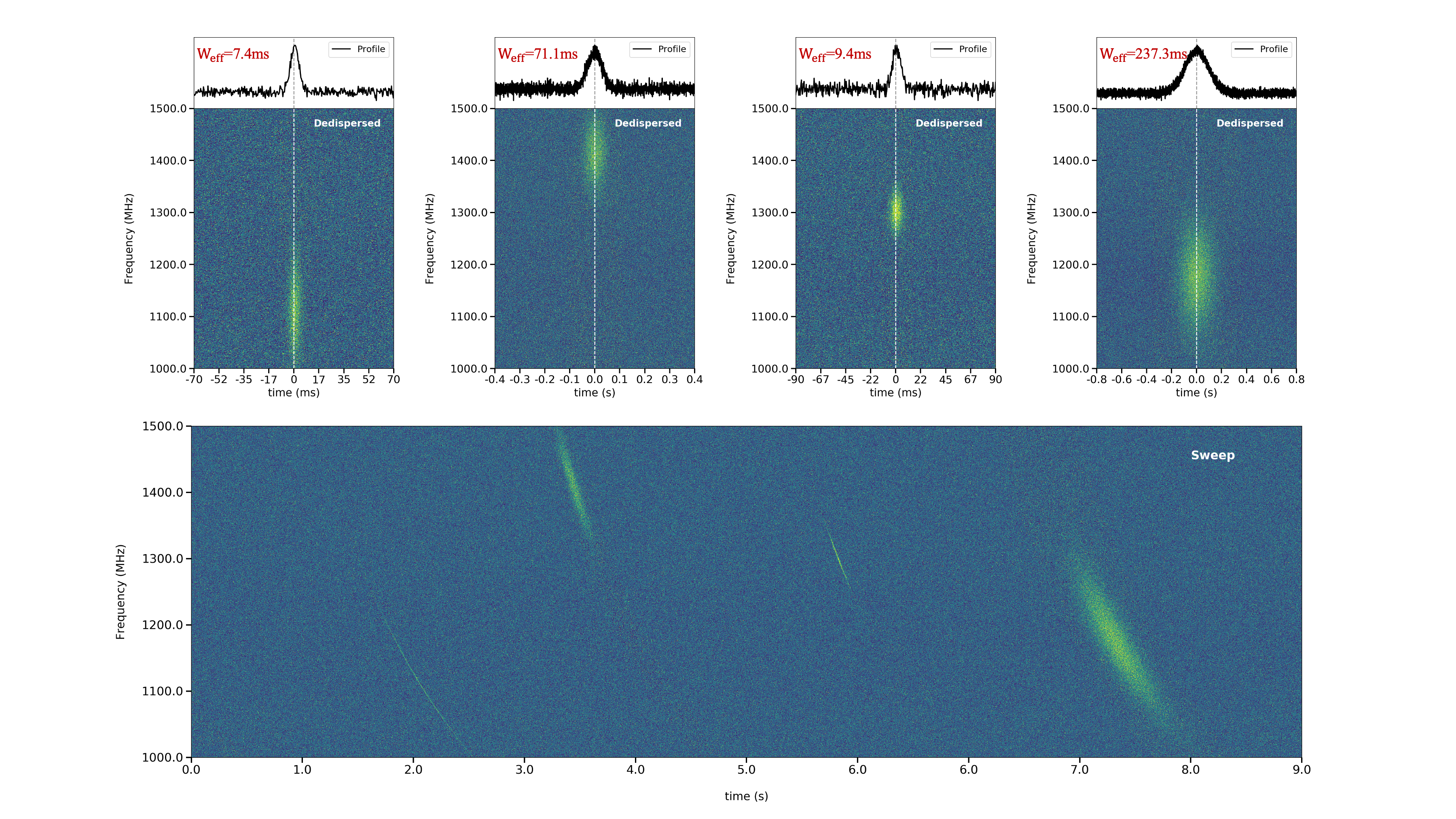}
	\caption{\textbf{Examples of typical simulated bursts.} The top four panels show dedispersed dynamic spectra of simulated bursts with \(W_{\text{eff}}\) = 7.4 ms, 71.1 ms, 9.4 ms, and 237.3 ms, respectively. Each panel includes the frequency-time waterfall plot and the corresponding burst profile above it. The bottom panel shows the simulated dispersed signals before dedispersion.}
	\label{fig-method-simu}
\end{figure*}

\begin{table}
  \centering
  \begin{tabularx}{\textwidth}{lXXXXXXXX} 
    \hline
    & Recall & Precision & Efficiency & Low BW Rec. & High TW Rec. & TW Smoothness & Low SNR Rec. \\
    \hline
    EDEN    & 0.730 & 0.204 & 2.827 & 0.364 & 0.425 & 0.346 & 0.353 \\
    Heimdall & 0.509 & 0.007 & 0.668 & 0.236 & 0.308 & 0.233 & 0.059 \\
    \hline
  \end{tabularx}
  \caption{Quantitative comparison between EDEN and Heimdall on simulated data.}
  \label{tab:quant_comparison}
\end{table}

\begin{figure}[ht]
\centering
{
    \begin{minipage}[b]{.3\linewidth}
        \centering
        \includegraphics[scale=0.28]{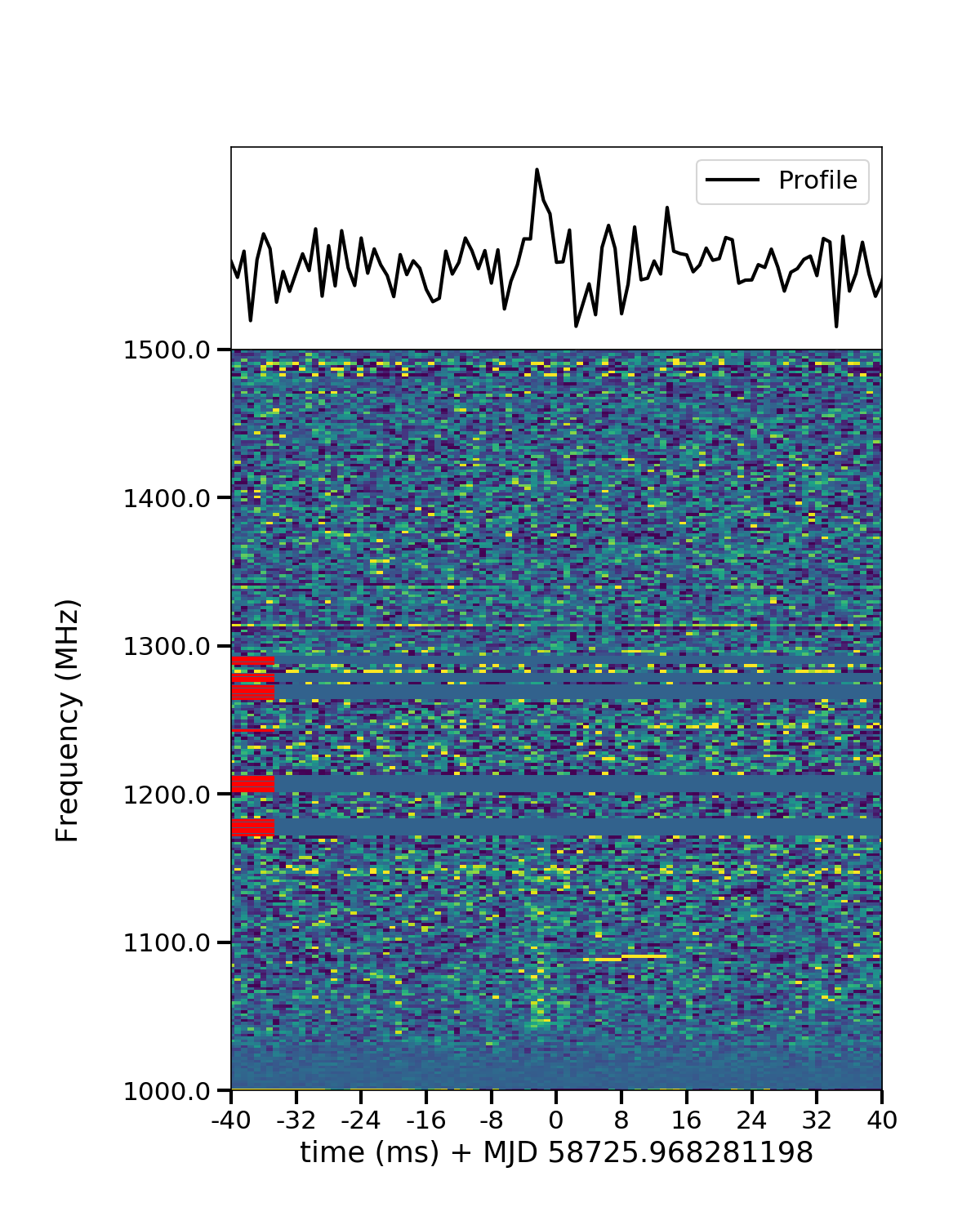}
    \end{minipage}
}
{
 	\begin{minipage}[b]{.3\linewidth}
        \centering
        \includegraphics[scale=0.28]{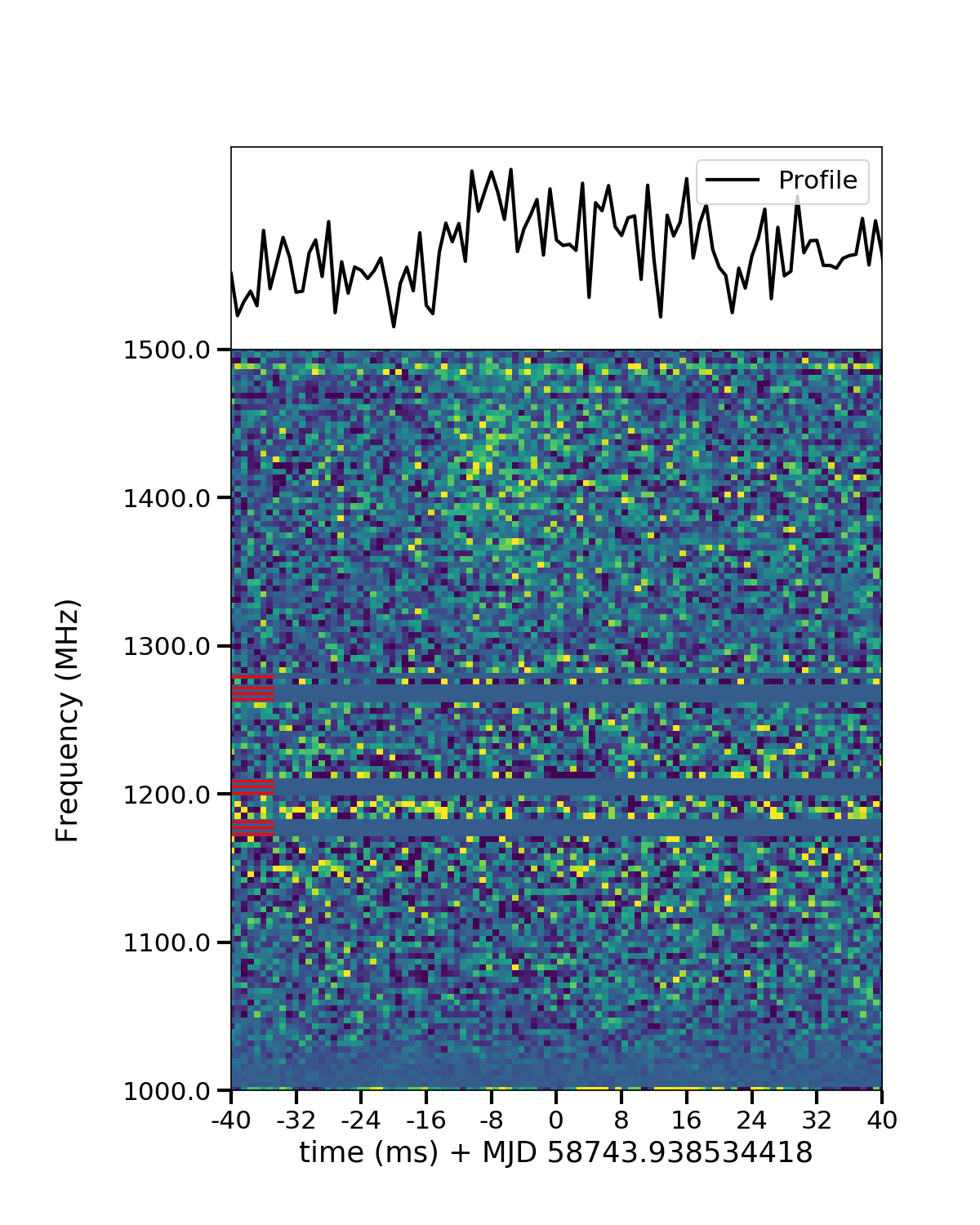}
    \end{minipage}
}
{
 	\begin{minipage}[b]{.3\linewidth}
        \centering
        \includegraphics[scale=0.28]{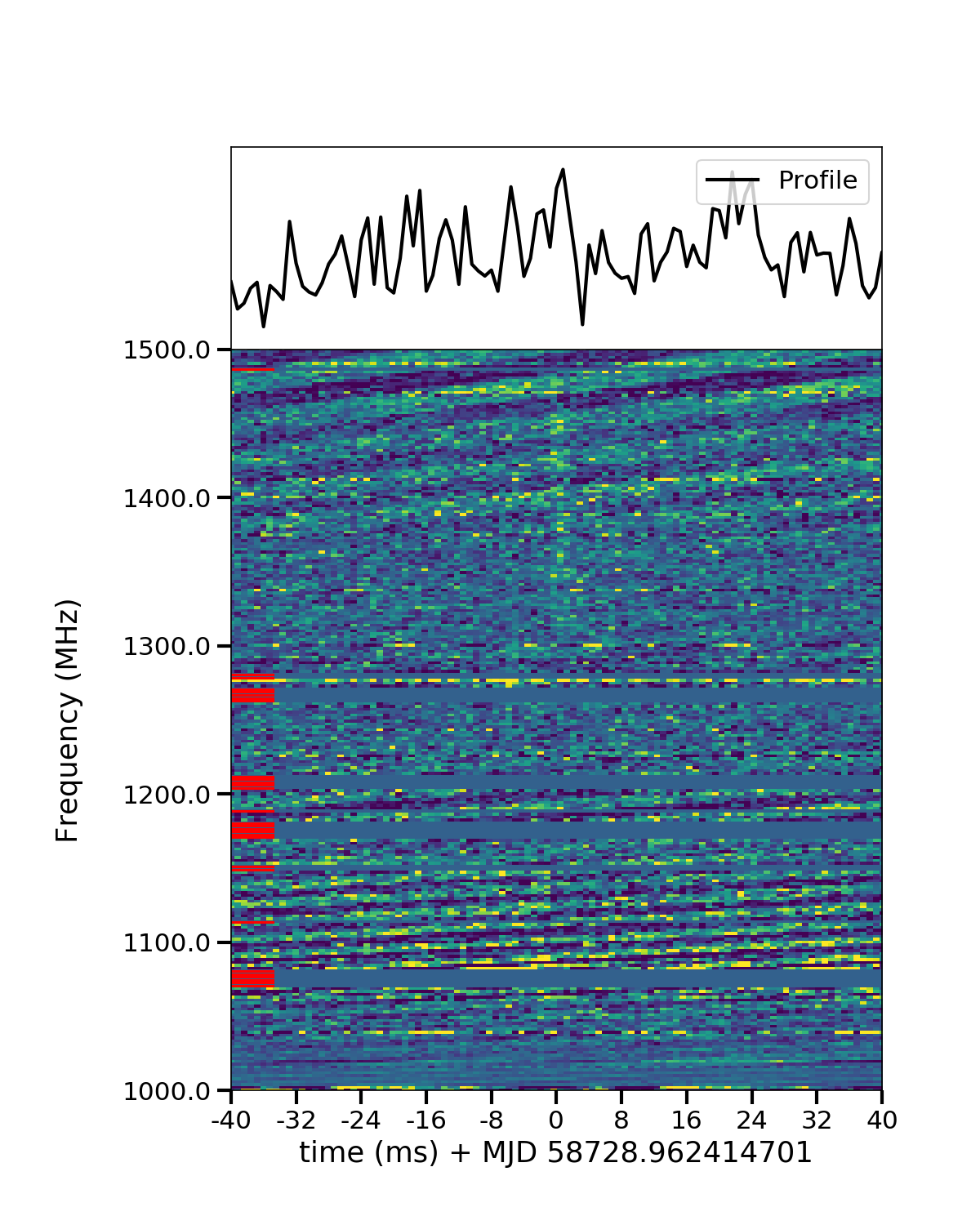}
    \end{minipage}
}
    \caption{\textbf{Examples of newly detected weak signals by EDEN.} Left: a typical example of temporal narrow bursts (\(W_{eff}=2.0\) ms) with MJD\(_{topo}\)=58725.968281198. Center: a typical example of weak and wide bursts (\(W_{eff}=32.6\) ms) with MJD\(_{topo}\)=58743.938534418. Right: a typical example of bursts in strong background RFI with MJD\(_{topo}\)=58728.962414701.}
    \label{fig:weak_signals}
\end{figure}

\clearpage
\bibliographystyle{naturemag}

\subsection{Data availability}
Observational properties of 5927 burst events of FRB 20121102A measured with FAST from August 2019 to October 2019 are summarized in the manuscript Supplementary Table2. Observational data are available from the FAST archive\footnote{http://fast.bao.ac.cn} and \url{https://doi.org/10.11922/sciencedb.01092}. Due to the large data volume for these observations, interested users are encouraged to contact the corresponding author to arrange the data transfer. Webpage for demonstrating the burst properties and dataset can be seen at \url{https://fast.cstcloud.cn/datavolume/10.1038.s41586-021-03878-5}.

\subsection{Code availability}
Computational programs for the FRB121102 burst analysis and observations reported here are available at \url{https://github.com/NAOC-pulsar/PeiWang-code}. Other standard data reduction packages are available at their respective websites:\\
PRESTO: \url{https://github.com/scottransom/presto}\\
DSPSR:  \url{http://dspsr.sourceforge.net}\\
PSRCHIVE:  \url{http://psrchive.sourceforge.net}

\clearpage
\section*{Supplementary Table}
\renewcommand{\baselinestretch}{1.0}
\selectfont
\noindent
\EXTTAB{tab:obslog}~1: Observational parameters of FRB 20121102A from Aug. 2019 to Oct. 2019. \\%
\EXTTAB{tab:bursttab}~2: Properties of 5927 bursts of FRB 20121102A measured with FAST from Aug. 2019 to Oct. 2019. \\%

\setcounter{figure}{0}
\setcounter{table}{0}
\captionsetup[table]{name={\bf Supplementary Table} }
\setlength{\tabcolsep}{2.5mm}{
\renewcommand\arraystretch{1.1}
\scriptsize
\begin{longtable}{c c c c c c c c c}%
\caption{Observational parameters of FRB 20121102A from Aug. 2019 to Oct. 2019.}
\\
\hline%
\hline%
Observational Date & MJD$_{start}$ & MJD$_{end}$ & Telescope & Frequency & Duration & Number of & Burst Rate \\%
(\textit{UT, YYYYMMDD})& & & & (GHz) & (hour) & Detections(\(^\star \text{Num}_{\text{tot}}\)) & (hour$^{-1}$) \\%
\hline%
\endhead%
\hline%
\endfoot%
\hline%
\endlastfoot%
20190830  & 58724.863883206 & 58725.000011231 &  &  &  3    & 625 (634) & 208.3 \\%
20190831  & 58725.862045706 & 58726.000011231 &  &  &  3    & 504 (509) & 168.0 \\%
20190901  & 58726.911633866 & 58727.104733243 &  &  &  4.5  & 496 (501) & 110.2 \\%
20190902  & 58727.873562373 & 58728.083344437 &  &  &  5    & 601 (606) & 120.2 \\%
20190903  & 58728.951098646 & 58729.083344564 &  &  &  3    & 397 (450) & 132.3 \\%
20190905  & 58730.863607106 & 58730.909732956 &  &  &  1    & 234 (235) & 234.0 \\%
20190906  & 58731.967746539 & 58732.013405627 &  &  &  1    & 0 & 0.0 \\%
20190907  & 58732.860435532 & 58732.993066290 &  &  &  1    & 495 (499) & 495.0 \\%
20190908  & 58733.893162326 & 58733.932291368 &  &  &  0.9  & 248 & 275.6 \\%
20190909  & 58734.913506863 & 58735.048621845 &  &  &  1    & 263 (265) & 263.0 \\%
20190911  & 58736.952874826 & 58737.006955179 &  &  &  1    & 204 (205) & 204.0 \\%
20190913  & 58738.954502396 & 58739.000010734 &  &  &  1    & 116 & 116.0 \\%
20190914  & 58739.994251192 & 58740.042024623 &  &  &  1    & 0 & 0.0 \\%
20190915  & 58740.986759884 & 58741.035184345 &  &  &  1    & 0 & 0.0 \\%
20190916  & 58741.862676262 & 58741.916677401 &  &  &  1    & 6 & 6.0 \\%
20190917  & 58742.864667130 & 58742.916677401 &  &  &  1    & 6 & 6.0 \\%
20190918  & 58743.912780752 & 58743.958679716 &  &  &  1    & 12 & 12.0 \\%
20190919  & 58744.856440058 & 58744.909916177 &  &  &  1    & 13 & 13.0 \\%
20190920  & 58745.913144433 & 58745.965288512 &  &  &  1    & 59 & 59.0 \\%
20190921  & 58746.834358426 & 58746.902788512 &  &  &  1    & 116 & 116.0 \\%
20190922  & 58747.834473206 & 58747.868066207 &  &  &  0.8  & 46 & 57.5 \\%
20190923  & 58748.903421134 & 58748.951735271 & FAST & 1.0 - 1.5  &  1    & 118 & 118.0 \\%
20190924  & 58749.844908414 & 58749.895844067 &  &  &  1    & 109 & 109.0 \\%
20190925  & 58750.829750104 & 58750.895844067 &  &  &  1    & 98 & 98.0 \\%
20190926  & 58751.874056447 & 58751.923621845 &  &  &  1    & 72 & 72.0 \\%
20190927  & 58752.825412454 & 58752.888899623 &  &  &  1    & 110 & 110.0 \\%
20190928  & 58753.918636389 & 58753.953245075 &  &  &  0.8  & 103 (104) & 128.8 \\%
20190929  & 58754.977865231 & 58755.022800086 &  &  &  1    & 153 & 153.0 \\%
20190930  & 58755.873914572 & 58755.898857975 &  &  &  0.5  & 60 & 120.0 \\%
20191001  & 58756.836106400 & 58756.881955179 &  &  &  1    & 221 & 221.0 \\%
20191002  & 58757.891142083 & 58757.937510734 &  &  &  1    & 135 & 135.0 \\%
20191003  & 58758.926598796 & 58758.979177401 &  &  &  1    & 126 & 126.0 \\%
20191004  & 58759.928775289 & 58759.979177401 &  &  &  1    & 74 & 74.0 \\%
20191005  & 58760.928776262 & 58760.979177401 &  &  &  1    & 0 & 0.0 \\%
20191006  & 58761.917797315 & 58761.965392679 &  &  &  1    & 0 & 0.0 \\%
20191007  & 58762.810287870 & 58762.855485188 &  &  &  1    & 3 & 3.0 \\%
20191008  & 58763.807343935 & 58763.847232873 &  &  &  1    & 5 & 5.0 \\%
20191009  & 58764.925915243 & 58764.979177401 &  &  &  1    & 5 & 5.0 \\%
20191010  & 58765.899767766 & 58765.948263049 &  &  &  1    & 0 & 0.0 \\%
20191011  & 58766.908249780 & 58766.958344067 &  &  &  1    & 8 & 8.0 \\%
20191012  & 58767.930755984 & 58767.979177401 &  &  &  1    & 7 & 7.0 \\%
20191013  & 58768.900363137 & 58768.944455179 &  &  &  1    & 8 & 8.0 \\%
20191014  & 58769.898923588 & 58769.944455179 &  &  &  1    & 0 & 0.0 \\%
20191015  & 58770.892862118 & 58770.944455179 &  &  &  1    & 3 & 3.0 \\%
20191016  & 58771.896752222 & 58771.937536677 &  &  &  1    & 0 & 0.0 \\%
20191017  & 58772.890455764 & 58772.937510734 &  &  &  1    & 11 & 11.0 \\%
20191018  & 58773.855838785 & 58773.916677401 &  &  &  1    & 3 & 3.0 \\%
20191019  & 58774.896024236 & 58774.942568604 & FAST & 1.0 - 1.5  &  1    & 2 & 2.0 \\%
20191020  & 58775.880451505 & 58775.923621845 &  &  &  1    & 0 & 0.0 \\%
20191021  & 58776.823957488 & 58776.875010734 &  &  &  1    & 52 & 52.0 \\%
20191025  & 58780.803006493 & 58780.854177401 &  &  &  1    & 0 & 0.0 \\%
20191030  & 58785.879255417 & 58785.902788970 &  &  &  1    & 0 & \ 0.0 %

\label{tab:obslog}
\end{longtable}}
\begin{tablenotes}
\item[\(\star\))] $\star$) Number of detections only contains the bursts used for calibration, and \(\text{Num}_{\text{tot}}\) in the blanket contains the total detections, including extremely weak bursts and bursts occurring within a strong interference background, which are unsuitable for calibration.
\end{tablenotes}
\clearpage

\setcounter{figure}{1}
\setcounter{table}{1}
\captionsetup[table]{name={\bf Supplementary Table} }
\setlength{\tabcolsep}{0.8mm}{
\renewcommand\arraystretch{2.0}
\scriptsize

\begin{longtable}{c c c c c c c c c c c}%
\caption{Properties of 5927 bursts of FRB 20121102A measured with FAST from Aug. 2019 to Oct. 2019.}
\\
\hline%
\hline%
Burst & MJD$^{a)}$  & DM$^{b)}$ & W$_{eff}^{c)}$ & Bandwidth$^{d)}$ & Cntrl freq. & Peak\ flux & Fluence & Energy$_{1.25 GHz}^{e)}$ & Err\_Energy$_{1.25GHz}^{f)}$ \\
\textit{ID}&(\textit{inf. }) & (\textit{pc\ cm$^{-3}$}) & (\textit{ms}) & (\textit{MHz}) & (\textit{MHz}) & (\textit{mJy}) & (\textit{mJy\ ms}) & (\textit{erg}) & (\textit{erg}) \\
\hline%
\endhead%
\hline%
\endfoot%
\hline%
\endlastfoot%

1 & 58724.877969807 & 563.8 $^{+1.1}_{-1.1}$ & 4.4 $^{+1.3}_{-1.3}$ & 160 $^{+52}_{-52}$ & 1386 & 21.48 $^{+0.38}_{-0.38}$ & 93.49 $^{+0.47}_{-0.47}$ & 1.06E+38 & 5.33E+35 \\
2 & 58724.878524804 & 563.4 $^{+1.3}_{-1.3}$ & 6.1 $^{+0.9}_{-0.9}$ & 74 $^{+50}_{-50}$ & 1106 & 12.83 $^{+0.22}_{-0.22}$ & 78.71 $^{+1.54}_{-1.54}$ & 8.43E+37 & 1.65E+36 \\
3 & 58724.878695930 & 563.0 $^{+1.4}_{-1.4}$ & 4.9 $^{+1.4}_{-1.4}$ & 82 $^{+41}_{-41}$ & 1381 & 19.06 $^{+0.33}_{-0.33}$ & 94.15 $^{+0.46}_{-0.46}$ & 1.06E+38 & 5.21E+35 \\
4 & 58724.878839835 & 564.2 $^{+0.8}_{-0.8}$ & 5.0 $^{+2.5}_{-2.5}$ & 96 $^{+9}_{-9}$ & 1387 & 11.73 $^{+0.20}_{-0.20}$ & 59.25 $^{+0.49}_{-0.49}$ & 6.70E+37 & 5.59E+35 \\
5 & 58724.879395083 & 563.3 $^{+1.3}_{-1.3}$ & 5.7 $^{+1.7}_{-1.7}$ & 94 $^{+3}_{-3}$ & 1367 & 13.91 $^{+0.23}_{-0.23}$ & 79.22 $^{+0.41}_{-0.41}$ & 8.95E+37 & 4.61E+35 \\
 &  &  & \(\dots\) & \(\dots\) & \(\dots\) & \(\dots\) & \(\dots\) &  & \\
5923 & 58776.870599613 & 563.5 $^{+1.4}_{-1.4}$ & 3.3 $^{+1.6}_{-1.6}$ & 80 $^{+19}_{-19}$ & 1441 & 3.73 $^{+0.03}_{-0.03}$ & 12.15 $^{+0.05}_{-0.05}$ & 1.37E+37 & 5.33E+34 \\
5924 & 58776.872681670 & 564.6 $^{+1.2}_{-1.2}$ & 3.9 $^{+0.7}_{-0.7}$ & 168 $^{+35}_{-35}$ & 1308 & 11.58 $^{+0.09}_{-0.09}$ & 44.68 $^{+0.06}_{-0.06}$ & 5.05E+37 & 6.81E+34 \\
5925 & 58776.873012298 & 564.6 $^{+0.9}_{-0.9}$ & 2.2 $^{+0.3}_{-0.3}$ & 52 $^{+12}_{-12}$ & 1396 & 9.33 $^{+1.16}_{-1.16}$ & 20.71 $^{+0.37}_{-0.37}$ & 2.22E+37 & 3.97E+35 \\
5926 & 58776.873652143 & 564.0 $^{+1.3}_{-1.3}$ & 8.0 $^{+1.2}_{-1.2}$ & 76 $^{+20}_{-20}$ & 1236 & 9.32 $^{+0.07}_{-0.07}$ & 74.40 $^{+0.68}_{-0.68}$ & 7.96E+37 & 7.24E+35 \\
5927 & 58776.873917995 & 564.9 $^{+0.1}_{-0.1}$ & 2.6 $^{+0.4}_{-0.4}$ & 401 $^{+14}_{-14}$ & 1287 & 46.05 $^{+0.37}_{-0.37}$ & 119.04 $^{+1.08}_{-1.08}$ & 1.27E+38 & 1.16E+36
\label{tab:bursttab}
\end{longtable}

\begin{tablenotes}
\item[a)] $a)$ The burst peak's arrival MJD modified to infinity frequency. \\%
\item[b)] $b)$ DM is obtained using \textit{DM-Power} algorithm.\\%
\item[c)] $c)$ The burst width \(W_{eff}\) is defined as the effective width of a rectangular burst that has the same area as the original burst, where the height of the rectangle is equal to the maximum value of the baseline-corrected spectrum.\\%
\item[d)] $d)$ Bandwidth is calculated separately by CDF fitting algorithm.\\%
\item[e)] $e)$ Full bandwidth (with the center frequency = 1.25 GHz) was involved when the energy of bursts is calculated.\\%
\item[f)] $f)$ The error of the isotropic equivalent energy.\\%

\end{tablenotes}

}\end{methods}
\end{document}